\documentclass[twocolumn]{aastex62}

\usepackage{lineno}

\usepackage{newtxtext,newtxmath}
\usepackage[T1]{fontenc}
\usepackage{graphicx}	
\usepackage{amsmath}	

\usepackage{lineno}
\usepackage{xcolor}
\definecolor{darkgreen}{rgb}{0.1, 0.6, 0.1}

\newcommand{\msun}{$\rm M_\odot$}

\begin{document}


\title[Stellar assembly in UDGs]{Large dark matter content and steep metallicity profile predicted for Ultra-Diffuse Galaxies formed in high-spin halos}
\shorttitle{ Stellar population in UDGs }
\shortauthors{J. A. Benavides et al.}

\author[0000-0003-1896-0424]{Jos\'e A. Benavides\href{https://orcid.org/0000-0003-1896-0424}{\includegraphics[scale=0.8]{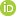}}}
\affiliation{Department of Physics and Astronomy, University of California, Riverside, CA, 92507, USA}

\author[0000-0002-3790-720X]{Laura V. Sales\href{https://orcid.org/0000-0002-3790-720X}{\includegraphics[scale=0.8]{images/orcid.png}}}
\affiliation{Department of Physics and Astronomy, University of California, Riverside, CA, 92507, USA}

\author[0000-0003-3055-6678]{Mario. G. Abadi\href{https://orcid.org/0000-0003-3055-6678}{\includegraphics[scale=0.8]{images/orcid.png}}}
\affiliation{Instituto de Astronom\'ia Te\'orica y Experimental, CONICET-UNC, Laprida 854, X5000BGR, C\'ordoba, Argentina}
\affiliation{Observatorio Astron\'omico de C\'ordoba, Universidad Nacional de C\'odoba, Laprida 854, X5000BGR, C\'ordoba, Argentina}

\author[0000-0001-8593-7692]{Mark Vogelsberger\href{https://orcid.org/0000-0001-8593-7692}{\includegraphics[scale=0.8]{images/orcid.png}}}
\affiliation{Department of Physics, Kavli Institute for Astrophysics and Space Research, Massachusetts Institute of Technology, Cambridge, MA 02139, USA}

\author[0000-0003-3816-7028]{Federico Marinacci\href{https://orcid.org/0000-0003-3816-7028}{\includegraphics[scale=0.8]{images/orcid.png}}}
\affiliation{Department of Physics \& Astronomy ``Augusto Righi'', University of Bologna, via Gobetti 93/2, 40129 Bologna, Italy}
\affiliation{INAF, Astrophysics and Space Science Observatory Bologna, Via P. Gobetti 93/3, 40129 Bologna, Italy}


\author[0000-0001-6950-1629]{Lars Hernquist\href{https://orcid.org/0000-0001-6950-1629}{\includegraphics[scale=0.8]{images/orcid.png}}}
\affiliation{Harvard-Smithsonian Center for Astrophysics, Cambridge, MA 02138, USA}

\begin{abstract}
We study the stellar properties of a sample of simulated ultra-diffuse galaxies (UDGs) with stellar mass $\rm{M_\star=10^{7.5} - 10^{9} ~ M_{\odot}}$, selected from the TNG50 simulation, where UDGs form mainly in high-spin dwarf-mass halos. We divide our sample into star-forming and quenched UDGs, finding good agreement with the stellar assembly history measured in observations. Star-forming UDGs and quenched UDGs with $\rm{M_\star \geq 10^8 ~ M_\odot}$ in our sample are particularly inefficient at forming stars, having $2$ - $10$ times less stellar mass than non-UDGs for the same virial mass halo. These results are consistent with recent mass inferences in UDG samples and suggest that the most inefficient UDGs arise from a late assembly of the dark matter mass followed by a stellar growth that is comparatively slower (for star-forming UDGs) or that was interrupted due to environmental removal of the gas (for quenched UDGs). Regardless of efficiency, UDGs are $60\%$ poorer in [Fe/H] than the population of non-UDGs at a fixed stellar mass, with the most extreme objects having metal content consistent with the simulated mass-metallicity relation at $z \sim 2$. Quenched UDGs stop their star formation in shorter timescales than non-UDGs of similar mass and are, as a consequence, alpha-enhanced with respect to non-UDGs. We identify metallicity profiles in UDGs as a potential avenue to distinguish between different formation paths for these galaxies, where gentle formation as a result of high-spin halos would present well-defined declining metallicity radial profiles while powerful-outflows or tidal stripping formation models would lead to flatter or constant metallicity as a function of radius due to the inherent mixing of stellar orbits.
\end{abstract}


\section{Introduction}\label{sec:intro}

Ultra-Diffuse Galaxies (UDGs) offer an opportunity to understand and test galaxy formation models within $\Lambda$CDM. UDGs are, by definition, extreme objects: they represent the lowest surface brightness end of the dwarf galaxy population. While limits are not well established and several selection criteria have been applied in the literature, UDGs typically refer to dwarf galaxies with stellar masses $\rm{ M_\star \sim 10^8~ M_\odot}$ and effective radius $\geq 1.5$ kpc \citep{vanDokkum2015a,vanDokkum2015b}. UDGs do not appear to be overall extended, as their outer light radius seems comparable to other galaxies with similar mass \citep{Chamba2022}. Instead, their light is distributed centrally less concentrated than non-UDG dwarfs, giving them their defining large half-light radius and low central surface brightnesses. The quest is then to understand whether such suppression of stars at the center of UDGs is a natural consequence of known galaxy-formation processes or if additional mechanisms are needed in order to explain the observed properties of UDGs.\\

Forming dwarf galaxies with extended half-light radius seems not too difficult to accommodate within the $\Lambda$CDM framework. Several mechanisms have been proposed that could lead to the formation of UDG-like objects, including large angular momentum content in the gas due to high-spin halos \citep{Amorisco2016, Rong2017,Liao2019,Benavides2023}, diffuse stellar distribution due to bursty star formation \citep{DiCintio2017, Chan2018} or mergers \citep{Wright2021}. In addition, for high-density environments, tidal stripping \citep{Carleton2019, Sales2020}, tidal heating \citep{Jiang2019b}, expansion due to a change of the inner potential driven by gas removal \citep{Safarzadeh2017} or stellar fading after quenching \citep{Tremmel2020} should also be at play.\\

However, the peculiarity of UDGs seems to extend beyond their light distribution. There is substantial evidence suggesting that at least a fraction of the observed UDGs show divergent kinematic \citep[e.g., ][]{Toloba2018,ManceraPina2019a,Danieli2019,Gannon2024}, metallicity \citep{Buzzo2024} and/or globular cluster content \citep[see e.g.,][]{Peng2016,vanDokkum2017,Janssens2022,Saifollahi2022,Danieli2022} when compared to other non-UDG dwarfs with similar mass. It is unclear that these additional trends can be explained by one (or a combination of) formation mechanisms described above. For completeness, it is also important to acknowledge that some of the samples or subsamples of UDGs in the literature are consistent with being simply the diffuse-end of the dwarf population \citep[e.g.][]{Beasley2016, Roman2017a, Conselice2018, Lee2020, Iodice2020, Gannon2021, Saifollahi2021, Iodice2023, Zoller2024}, but questions remain on the nature of the most extreme objects \citep[e.g., ][]{Chilingarian2019, vanDokkum2019a_DF4,Danieli2022,Toloba2023,Doppel2024,ForbeGannon2024,Montes2024}. A good way to summarise the state of affairs is perhaps to postulate that UDGs in observations show too wide a range of properties to be explained by the suggested theoretical mechanisms to form these diffuse galaxies.\\

Of particular interest is the subpopulation of UDGs consistent with being ``failed galaxies" \citep{vanDokkum2015a,Forbes2020}, UDGs that inhabit massive dark matter halos given their stellar mass. Their overly-massive halos could explain the larger-than-normal number of globular clusters (GCs) observed in some UDGs while accommodating some of the large velocity dispersion measurements $\sigma \geq 40$ km/s \citep[e.g., ][]{Toloba2023}. Failed galaxies are also expected to have ancient stellar populations, low metallicities and to be alpha-enhanced as a result of an efficient interruption of their expected stellar mass build-up at early redshifts. Support for the existence of such UDG class has recently been presented in \citep{FerreMateu2023,Buzzo2024}, although it still awaits confirmation from kinematical measurements of halo mass in samples that include spectroscopic data \citep{Gannon2024}.\\ 

We note that the term ``failed galaxy" is also used some times in the literature to refer to UDGs hosted by dwarf-mass dark matter halos ($\rm{ M_{vir} < 10^{12} ~ M_\odot}$) as long as the star formation is more inefficient than in other dwarfs, resulting on old stellar populations, low metallicities and more than $\sim 20$ GCs \citep{FerreMateu2023,Gannon2024}. 
Observationally, with the exception of a few objects in the \citet{Toloba2023} sample, most UDGs with available mass determinations seem in agreement with a dwarf-halo mass range. This includes mass inferences based on scaling relations \citep{Zaritsky2023}, weak lensing \citep{Sifon2018}, globular cluster counts \citep{Amorisco2018,Lim2018,Saifollahi2022} and spectroscopic measurements from stars or GCs \citep[see data compilation in ][]{Gannon2024}. Many theoretical models studying the population of UDGs also agree with such mass range \citep{Amorisco2016, Rong2017, Sales2020, Wright2021, Kravtsov2024}. The key question to be answered is what turns these objects potentially less efficient to form stars compared to non-UDG dwarfs and do such processes leave an impact on other stellar properties such as metallicity or metallicity profiles.\\

In this work we investigate these questions in more detail for the formation scenario where UDGs are hosted in preferentially large spin-halos. Our sample is drawn from the Illustris TNG50 simulation and was introduced before in \citet{Benavides2021,Benavides2023}. We now extend the study in light of new observational constraints on two key aspects: star-formation efficiency and the identification of possible avenues to form metal-poor and/or alpha-enhanced objects within the UDG population. This paper is organized as follows. In Sec.~\ref{sec:sims} we briefly describe the simulation and some details of our sample of UDGs. In Sec.~\ref{sec:eff} and ~\ref{sec:metals} we analyze their star-formation efficiency, assembly history and metallicity properties. We summarize our main findings in Sec.~\ref{sec:concl}.\\

\section{Simulations and method}
\label{sec:sims}

We use the highest resolution TNG50-1 (TNG50 hereafter) cosmological hydrodynamical simulation \citep{Pillepich2018a, Pillepich2018b,Nelson2018,Naiman2018,Marinacci2018, Springel2018,Weinberger2018, Pillepich2019, Nelson2019TNG}. In TNG50 we can follow and analyze the evolution of galaxies within a $\sim 50$ Mpc on-a-side cosmological box using the {\sc arepo} code \citep{Springel2010}. The simulation assumes a set of cosmological parameters consistent with the \citet{PlankColaboration2016} measurements\footnote{cosmological constant $ \Omega_{\Lambda} = 0.6911$, matter content (dark matter + baryons) $\Omega_m = \Omega_{dm} + \Omega_b = 0.3089 $, $ \Omega_b = 0.0486 $, Hubble constant $\rm{ H_0 = 100 \, h \, km \, s^{-1} \, Mpc^{-1} }$, $ h = 0.6774 $, $\sigma_8 = 0.8159 $ and spectral index $ n_s = 0.9667 $.}. The mass resolution for dark matter particles is $\rm{m_{drk}} = 4.5 \times 10^5, \rm{M_{\odot}}$ and baryonic elements (cells for gas, particle for stars) are $\rm{m_{bar}} \sim 8.5 \times 10^4, \rm{M_{\odot}}$. Gravitational softening is the same for dark matter and stars and is $\epsilon^{z=0}_{DM, \star} = 0.29$ kpc for these collisionless components, but can be significantly smaller for the gas cells in high density regions, which is typically resolved with a minimum softening $50 ~ \rm{pc \, h^{-1}}$.\\

The baryonic modeling in TNG50 builds upon the groundwork laid by the previous Illustris project \citep{Vogelsberger2013, Vogelsberger2014a}, with adjustments made to the stellar and AGN feedback mechanisms as detailed primarily in \cite{Pillepich2018a} and \cite{Weinberger2017}, respectively. In summary, gas cooling is enabled down to a temperature of $T = 10^4 ~\rm K$, guided by locally computed cooling and heating rates which account for density, redshift, and metallicity. When gas cells above that reach a density threshold of $n_H \simeq 0.1$ $\rm cm^{-3}$ they stochastically form stars according to the Kennicutt–Schmidt relation \citep{Kennicutt1998}. In this process, the entire cell is converted to a star particle that inherits the cell's mass, momentum, and metallicity. Stellar particles are born assuming a Chabrier initial mass function \citep{Chabrier2003} and their stellar evolution follows the prescription described in \citep{Pillepich2018a}, where the stellar populations evolve and return mass and metals to the interstellar medium (ISM) through supernovae Type Ia (SNIa) and Type II (SNII). \\

In the TNG50 simulation, all stars contribute to the chemical enrichment of their surroundings via stellar winds and supernovae. To achieve this, metal-rich gas is injected into all cells within a sphere surrounding the star particle. Essentially, newly ejected metals from a star particle mix into the surrounding gas. During this process, the mass of the star particle is proportionally reduced, while its metallicity remains fixed. It is worth noting that the maximum fraction of mass that star particles, representing typical stellar populations, can lose is about $50 \%$, ensuring that no star particle is completely destroyed in this process. The simulation is initialized with a metallicity mass fraction $Z = 10^{-10}$ for all elements except hydrogen and helium \citep{Pillepich2018a}. In total, the TNG50 simulation provides information on the production and subsequent evolution of nine elements (H, He, C, N, O, Ne, Mg, Si, and Fe). More details about this kind of cosmological simulations are presented in \citet{Vogelsberger2014nat, Vogelsberger2020nat}.\\

\begin{figure*}
\centering
\includegraphics[width=\columnwidth]{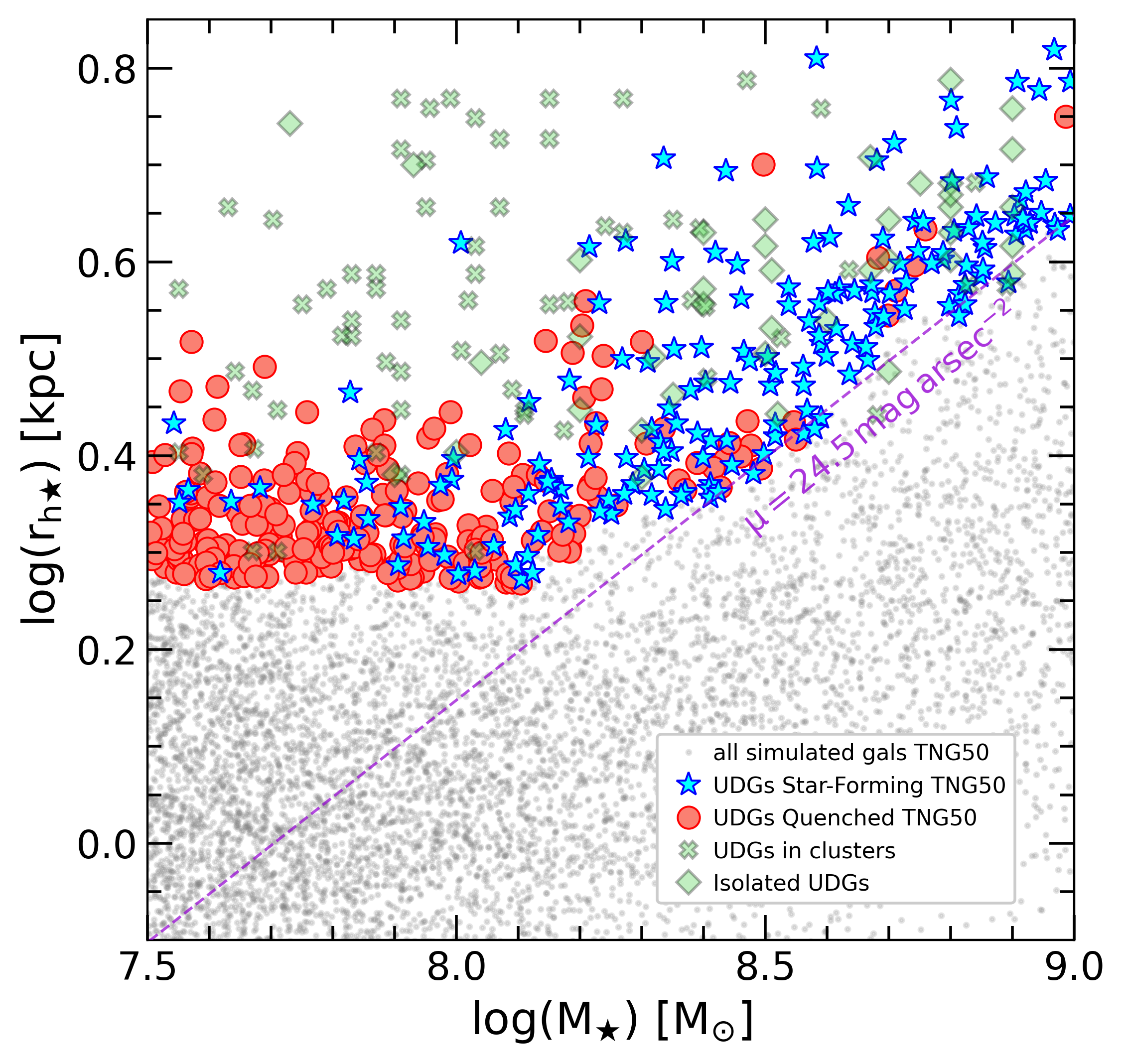}
\includegraphics[width=0.973\columnwidth]{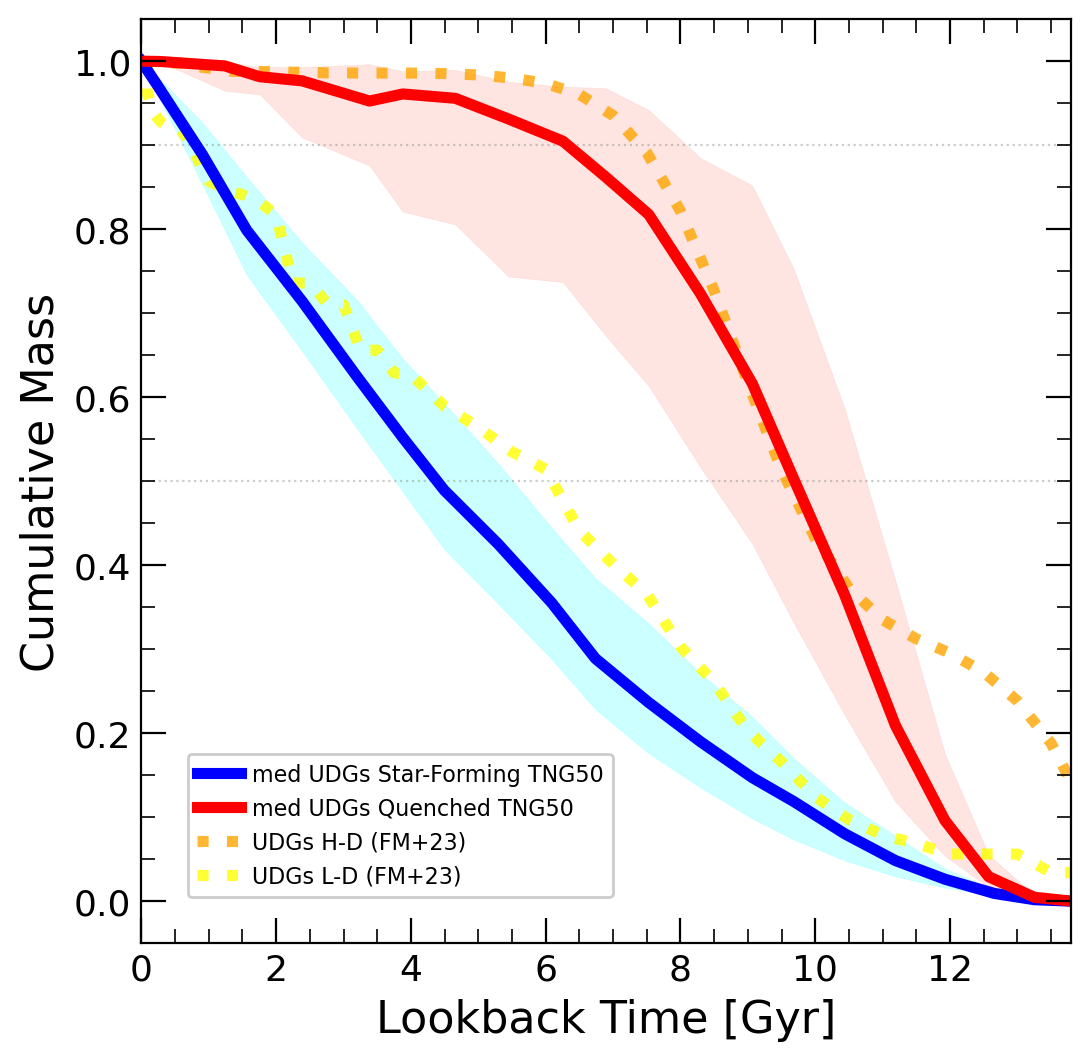}
\caption{\textit{Left}: Stellar mass vs. 3D stellar half-mass radius relation. All simulated galaxies in TNG50 are presented in grey dots, as well as the catalog of simulated UDGs of \citet{Benavides2021,Benavides2023}, separated into star-forming (blue stars) and quenched (red circles). For comparison with observations, we include a compilation of UDGs in clusters \citep{vanDokkum2015a, Lim2020, FerreMateu2023} and in isolation \citep{Leisman2017, ManceraPina2020, Rong2020b} with green symbols. \textit{Right}: Stellar mass assembly history of the population of simulated UDGs (solid lines) separated into star-forming (blue) and quenched (red), as in the left panel. The profiles correspond to the medians of each sample and the shaded regions to the 25th and 75th percentiles. The population of quenched UDGs assembled $50\%$ ($t_{50}$) of their stellar mass $\sim 10$ Gyr ago, while the star-forming galaxies just $\sim 4$ Gyr ago. The simulated assembly histories seem in reasonably good agreement with quenched UDGs from low and high-density regions from \citet{FerreMateu2023} (orange/yellow dotted lines).}
\label{fig:sample}
\end{figure*}

We use the friends-of-friends \citep[FoF]{Davis1985} and {\sc subfind} \citep{Springel2001, Dolag2009} group finding algorithms to identify bound structures (halos and subhalos) in the simulation. Object evolution over time is tracked using the SubLink merger trees \citep{RodriguezGomez2015}. Viral quantities, such as mass, radius, and velocity ($\rm{M_{200}}$, $\rm{r_{200}}$, and $\rm{V_{200}}$ respectively) are determined by the radius where the average density equals $200$ times the critical density of the Universe ($\rm{\rho_c = 3 H^2 / 8 \pi G}$). In the volume of the TNG50 simulation, we can analyze various environments at the same time, since galactic MW-like halos ($\rm{M_{200} \gtrsim 10^{12} ~ M_{\odot}}$) to clusters ($\rm{M_{200} \gtrsim 10^{14} ~ M_{\odot}}$) are present, as are thousands of dwarf galaxy - mass objects in the field.\\

We use interchangeably the terms {\it central} or {\it field} galaxy to refer to galaxies that are at the minimum of the gravitational potential of their FoF group and {\it satellite} to refer to anything associated with an FoF that is not central. Broadly, we will assume that central galaxies reside in the field, while satellites might belong to a galaxy-, group- or cluster- environment according to the virial mass of their host FoF halo. In this work, we define the infall time ($t_{\rm inf}$) as the snapshot before the first time a halo is identified as a satellite (or first infall). Galactic properties, such as stellar or gas mass, and star formation rates, are calculated based on all particles located within the ``galaxy radius'' ($\rm{r_{gal}}$). This radius is defined as twice the half-mass radius of the stars, denoted as $\rm{r_{h,\star}}$.

\subsection{Catalog of simulated UDGs in TNG50}
\label{ssec:sample}

For this work, we use the same catalog with 436 simulated UDGs (centrals + satellites) introduced in \citet{Benavides2021, Benavides2023}. UDGs are defined to be the most extreme outliers (top $5\%$) of the stellar mass - size relation predicted by the simulation in the stellar mass range $\rm{ M_{\star} = [10^{7.5}, 10^9] ~ M_{\odot}}$. At the mass resolution of TNG50, our lower limit corresponds to objects with $\sim 300$ stellar particles, ensuring robust determination of their sizes. Note that numerical artifacts, such as the spurious heating due to the mass difference between baryonic and dark matter particles, may be unavoidable for this low-mass scale, inflating the typical sizes of low-mass objects \citep{Ludlow2023}. We argue here that our strategy of selecting the UDGs not based on a fixed size cut-off (as usually assumed on other theoretical or observational samples) but instead as the most extreme outliers of the typical mass-size relation of the simulation manages to identify those objects that for physical reasons are more extended than the ``normal'' population of galaxies.\\  

The left panel of Fig.~\ref{fig:sample} introduces our sample in the stellar mass-size relation and compares it to several observational sets of UDGs from the literature, both in clusters \citep{vanDokkum2015a, Lim2020, FerreMateu2023} and in isolation \citep{Leisman2017, ManceraPina2020, Rong2020b} with green symbols. We use the 3D half-mass radius $\rm{r_{h,\star}}$ as a proxy for size and adopt the relation $\rm{r_{h,\star} = 4/3 R_e}$ to scale the projected half-light radius in observational samples \citep{Hernquist1990}. The median of the simulated sample is indicated with a thick solid black line. The bound that defines the UDG sample is shown in the dotted black line, which indicates the top $5\%$ most-extended galaxies at a given $\rm{M_{\star}}$. UDGs are color-coded according to whether they are star-forming (blue) or quenched (red) at the present day in the simulation, using as an indicator the $z=0$ specific star formation rate $\rm{sSFR = 10^{-11} ~ yr^{-1}}$ as a threshold, following the criteria adopted by \citet{Wetzel2012}.\\

\begin{figure*}
\centering
\includegraphics[width=1.0\textwidth]{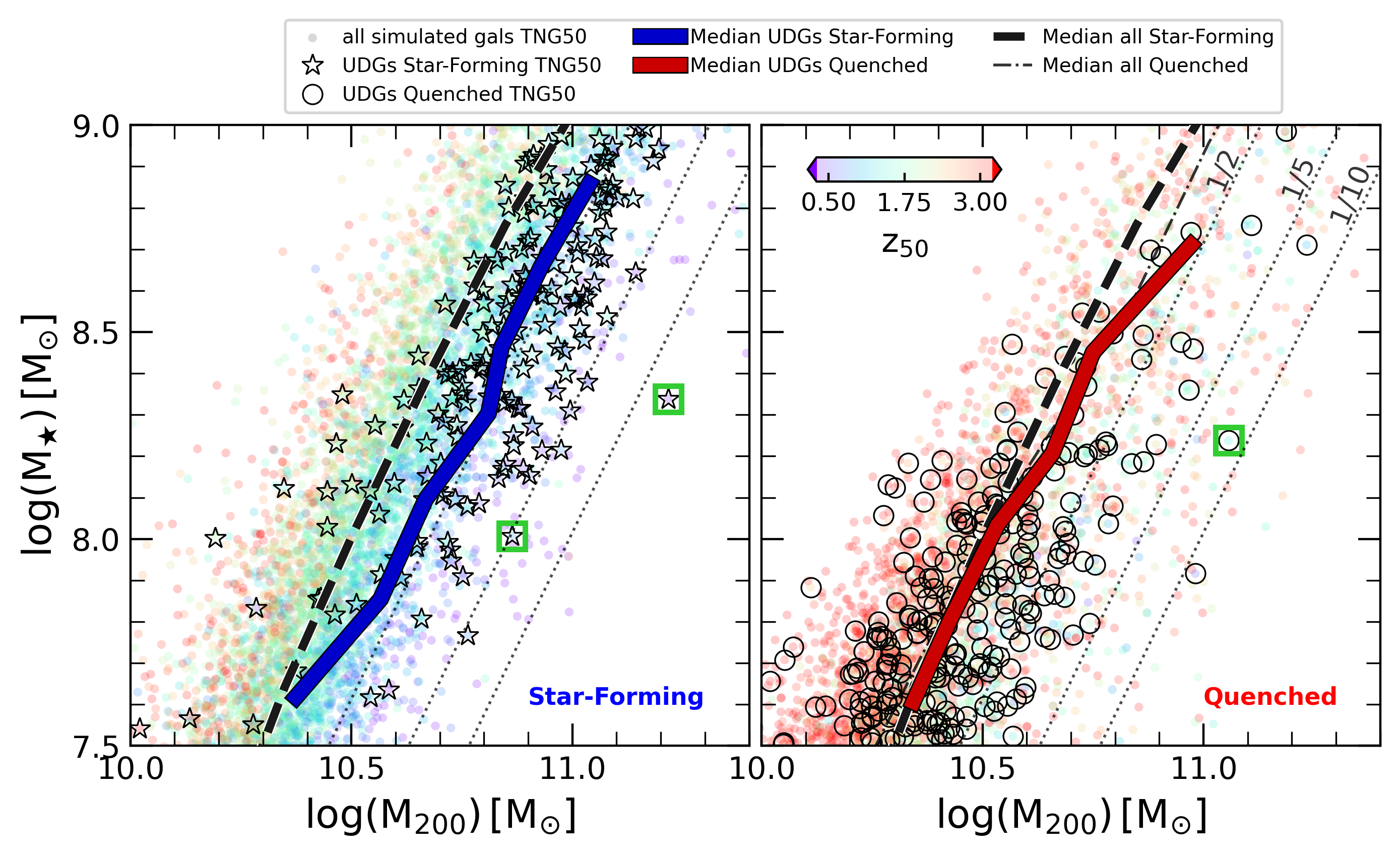}
\caption{Stellar mass as a function of halo virial mass for all simulated galaxies from TNG50. The color code indicates the redshift when the dark matter halo of each galaxy reached $50\%$ ($z_{\rm 50}$) of its virial mass today (for satellites the maximum virial mass as centrals is used instead). The population of galaxies has been separated into star-forming (left panel) and quenched (right panel), with UDGs highlighted in open stars and circles, respectively. The median of all star-forming galaxies is indicated with a dashed black line in both panels and the individual median for star-forming and quenched UDGs with solid blue and red lines, respectively. The median of all quenched galaxies is indicated with a thin dash-dotted black line (only) in the right panel. Thin dotted lines correspond to multiplying the median of all star-forming dwarfs (long dashed line) by factors $0.5$, $0.2$, and $0.1$ to show different levels of inefficiency at forming stars. Star-forming UDGs and massive quenched UDGs with $\rm{M_\star \geq 10^{8.2} ~ M_{\odot}} $ populate their halos with systematically less stellar mass than non-UDGs. The bias disappears for lower mass UDGs, whose median $\rm{M_\star - M_{200}}$ relation (thick solid red line) traces well the median of non-UDGs dwarfs. Note the abundance of UDGs forming $2$-$5$ fewer stars than non-UDGs with similar halo mass. The green unfilled squares highlight three individual examples presented in Figure~\ref{fig:examples}.}
\label{fig:mhalo_mstr}
\end{figure*}

As presented in detail in \citet{Benavides2023}, there are expected correlations between the stellar population of our simulated UDGs and their environment, with central or field galaxies being typically star-forming and gas-rich today and satellite dwarfs being quiescent. Note, however, that this is not always strictly true: $16\%$ of the quiescent UDGs are actually central galaxies in low-density environments today, which are all associated with backsplash orbits in the past as discussed in \citet{Benavides2021}. Similarly, $26\%$ of our star-forming sample is composed of satellite galaxies, which have been accreted only recently and still retain all or some of their gas. Environmental effects in our simulations include evolution within a wide range of host halos ranging from galaxy-like halos to medium-mass galaxy clusters (the most massive cluster in the simulation has $\rm{M_{200} \sim 2 \times 10^{14} ~ \rm{M_\odot}}$).\\

The right panel of Fig.~\ref{fig:sample} shows the typical stellar assembly histories of our simulated star-forming (blue) and quiescent (red) UDGs. Star-forming UDGs assemble more steadily over time and with a significant delay compared to the population of quenched galaxies. This is clearly shown by the steep slope of the solid red line, indicating that most of the stellar mass in quenched UDGs is attained in a short timescale before a lookback time $\sim 6$ Gyr and a stagnation in $M_\star$ afterwards. We can quantify these differences by calculating the time in which star-forming and quenched UDGs assemble $50\%$ of their final mass, $t_{50}$, for which we find $9.6\pm 1.5$ and $4.8\pm 1.5$ for star-forming and quenched UDGs, respectively. We have independently checked that taking all star-forming or quenched galaxies as a whole, independent of their size (UDGs and non-UDGs), would retrieve comparable assembly histories (red and blue solid lines), making the assembly of UDGs not particularly different than any other dwarf galaxy in the simulation. Encouragingly, the assembly history of our star-forming and quenched UDGs compares well with observational constraints for low-density and high-density regions, respectively, presented in \citet{FerreMateu2023} (dotted lines). The good agreement between our star-forming sample and the low-density quenched UDGs in \citet{FerreMateu2023} may also indicate that they have only very recently become quenched.\\

\section{The efficiency of star formation in UDGs}
\label{sec:eff}

Constraining the dark matter halo mass of UDGs has been a priority in understanding their formation mechanism. There seems to be a consensus that for some fraction of the UDGs the dark matter halo inferred seems in line with that expected of normal dwarf galaxies. However, observationally, the hypothesis that at least some of them may live in overly-massive halos comes from a variety of probes, including a large number of GCs \citep{Peng2016, Lim2020, Buzzo2024, ForbeGannon2024}, large velocity dispersion from their associated GCs \citep{Toloba2023} or stellar velocities \citep{Forbes2020,Gannon2022} and, more recently, from more indirect probes such as their low metallicity \citep[e.g.,][]{Kadowaki2017, FerreMateu2018, FerreMateu2023,Gu2018, RuizLara2018} and halo mass inferred through their position in galaxy scaling relations \citep{Zaritsky2023}.\\

\begin{figure*}
\centering
\includegraphics[width=\columnwidth]{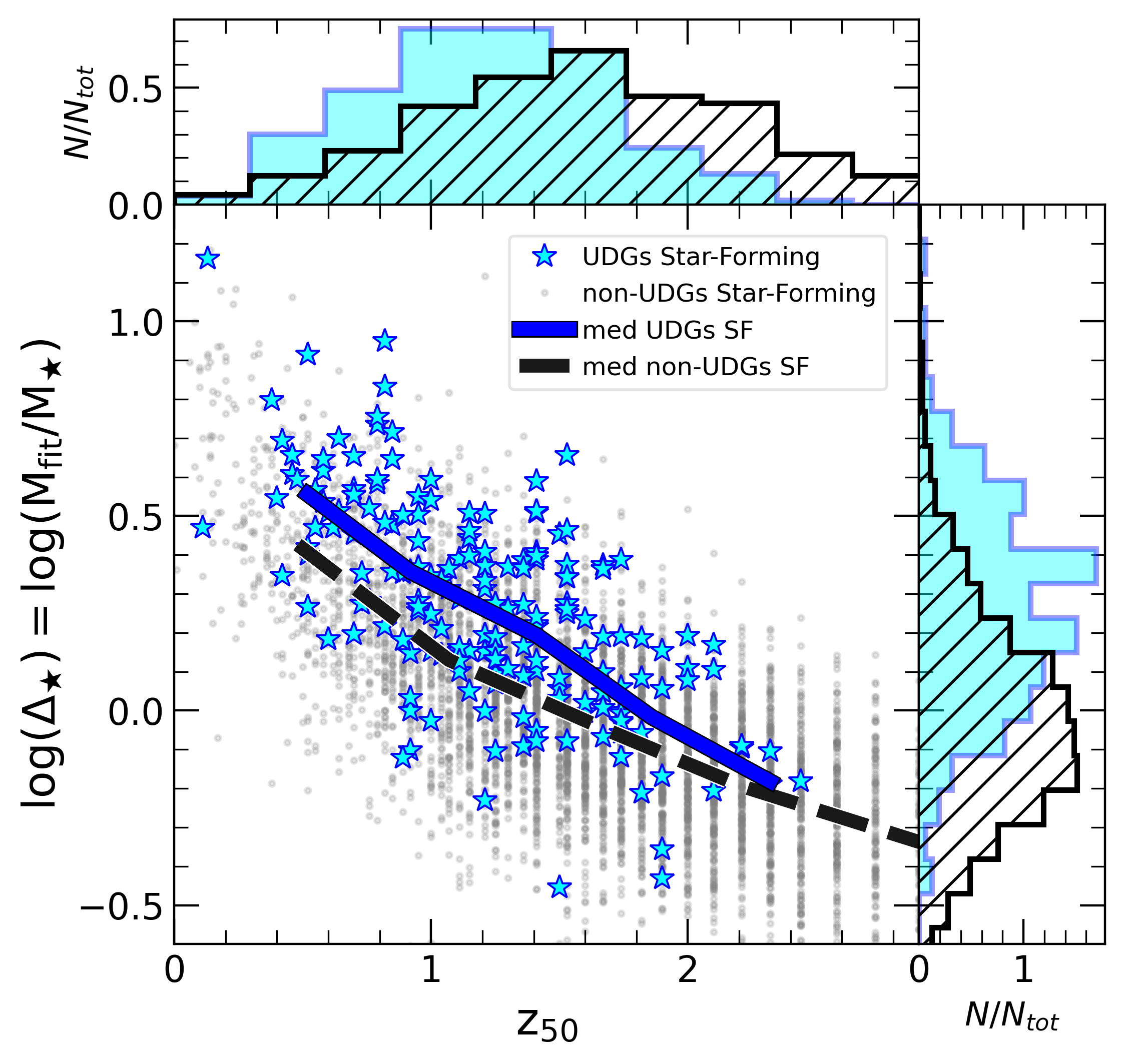}
\includegraphics[width=\columnwidth]{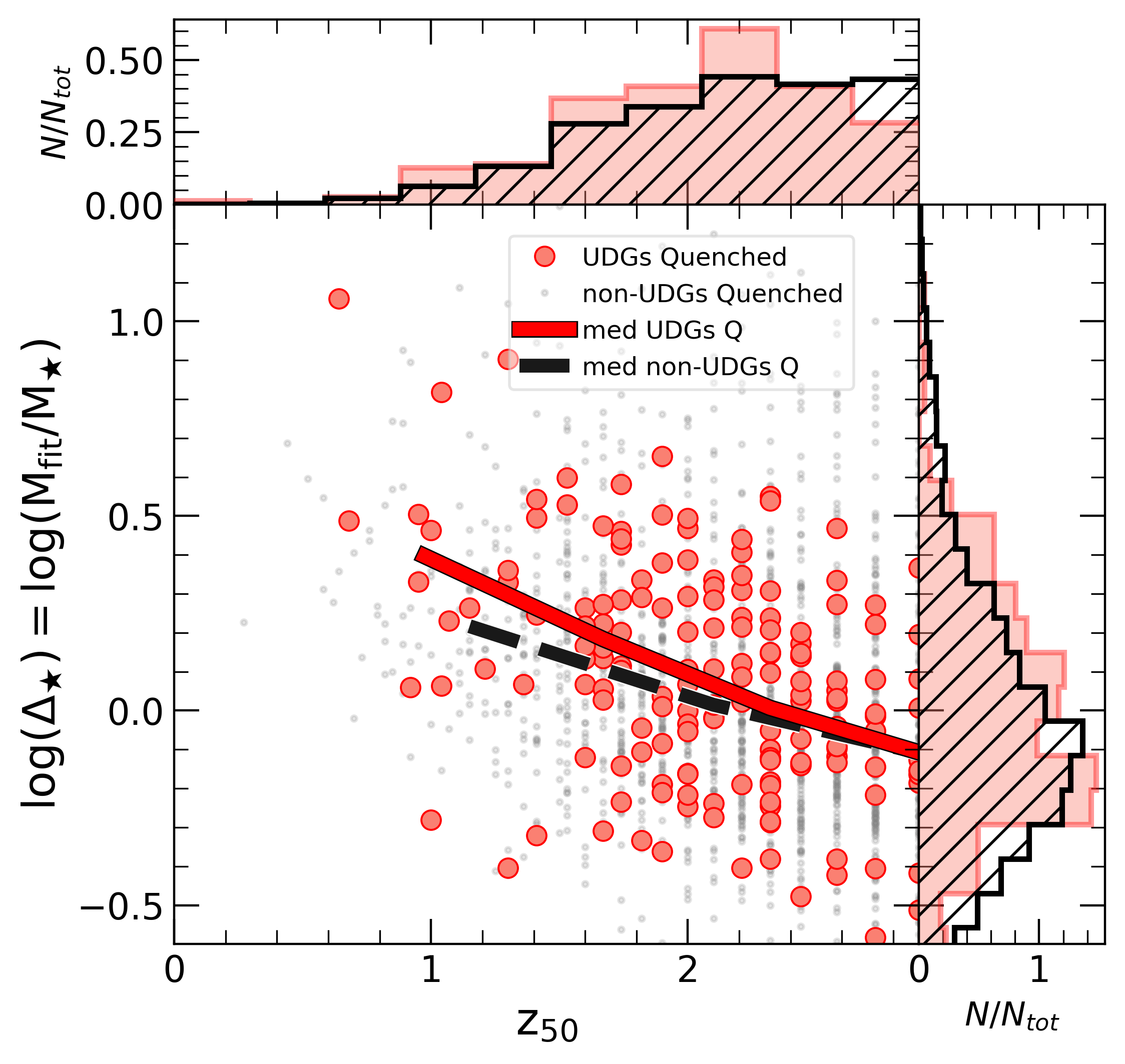}
\caption{Inefficiency of star formation $\Delta_\star$ as a function of the redshift when each dark matter halo assembled $50\%$ ($z_{50}$) of its final virial mass. $\Delta_\star$ is the ratio between the expected stellar mass($\rm{M_{ fit}}$) and their true stellar mass, where $\rm{M_{fit}}$ refers to 
 the median of $\rm{M_\star}$ expected at a given $\rm{M_{200}}$ from the non-UDG sample (thick black dashed curve in Fig.~\ref{fig:mhalo_mstr}). Star-forming galaxies are shown on the left panel and quenched on the right. Blue stars and red circles highlight UDGs and grey dots show non-UDGs galaxies. The median for each population is shown in dashed black (non-UDGs), blue or red solid line for UDGs star-forming or quenched, respectively. Halos that assemble later show more stellar mass suppression (larger $\Delta_\star$) for their halo mass. }
\label{fig:delta_m*}
\end{figure*}

In \citet{Benavides2023} we have already established that all UDGs formed in TNG50 lived in dark matter halos consistent with dwarf galaxies ($\rm{M_{200} \leq 10^{11.2} ~ \rm{M_\odot}}$). However, the hypothesis under scrutiny is that UDGs may live in dwarf-mass halos but were substantially less efficient to form the typical amount of stars, resulting on overly-massive objects in dark matter compared to their baryonic content. Because the assembly of galaxies that are star-forming or quenched can be quite different from each other, we will divide our sample into these two categories assuming that all galaxies with a specific star formation rate $\rm{sSFR< 10^{-11} ~ M_\odot ~ yr^{-1} }$ are quenched \citep{Wetzel2012}. The large majority ($\sim 73\%$) of star-forming galaxies (with $\rm{M_\star \in [10^{7.5}-10^9] ~M_\odot}$) are central or field galaxies while the majority of quenched galaxies are satellites ($\sim 88\%$). However, we choose to favor the star-forming vs. quenched classification instead of central vs. satellite since the division based on star formation activity is more easily comparable one-on-one to observations.\\

We investigate in Fig.~\ref{fig:mhalo_mstr} the relation between the stellar mass and halo mass in our sample, divided in star-forming (left panel) and quenched (right panel). Because $\rm{M_{200}}$ is ill-defined for satellites, for those cases (more common among quenched objects) we show their present-day stellar mass, $\rm{M_\star}$, but their halo mass $\rm{M_{200}}$ is defined as the maximum virial mass before they became a satellite. The left panel shows something quite interesting: star-forming UDGs (highlighted with dark starred symbols) are clearly less efficient at forming stars than non-UDGs. The median relation of all galaxies in TNG50 is marked with a thick black dashed line, indicating the median $\rm{M_\star}$ for a halo of a given mass. Star-forming UDGs lie below such relation on average by a factor $\sim 2$ (solid blue line is the corresponding median for UDGs) but some of our simulated UDGs can scatter further down reaching factors $\sim 3$ - $10$ fewer stars at fixed $\rm{M_{200}}$ than non-UDGs. This can be seen by comparison to the thin solid black lines, which show the median stellar mass - halo mass relation of all star-forming galaxies shifted by factors of $2$, $5$, and $10$ downwards in stellar mass. Our most extreme UDG example is $\sim 14$ times less massive in stars than what is expected from the median relation. The bias seems larger for the more massive end of our sample and may disappear for closer to $\rm{M_\star \sim 10^8 ~ M_\odot}$.\\

We trace the origin of this inefficiency for star-forming UDGs to their assembly time. All objects are color-coded by their formation time, defined here as the redshift when their virial mass reaches $50\%$ of their final value. Late-forming objects are located systematically lower in $\rm{M_\star}$ at a fixed halo mass, independent of whether they are UDGs or non-UDGs. By selecting extended stellar systems, we seem to be biasing the results to later forming halos which statistically show higher spins than those with earlier assembly (see Fig.~\ref{fig:app_lambda_z50} in the Appendix). Note that this is in good agreement with findings from UDGs in semi-analytical models as well \citep{Rong2017}. The late assembly of the dark matter in our systems seems to be accompanied by a slower build up of the stellar mass, resulting on dwarfs with an overly-massive dark matter content compared to their $\rm{M_\star}$. We have explicitly checked that UDGs follow the main trend in $\rm{M_{bar} - M_{200}}$ relation along with non-UDGs, where $\rm{M_{ bar}}$ refers to the gas and stellar content within the radius of the galaxy. It is only in terms of their stellar content (and not their total baryonic content) that star-forming UDGs are ``inefficient''.\\

\begin{figure*}
\centering
\includegraphics[width=0.325\textwidth]{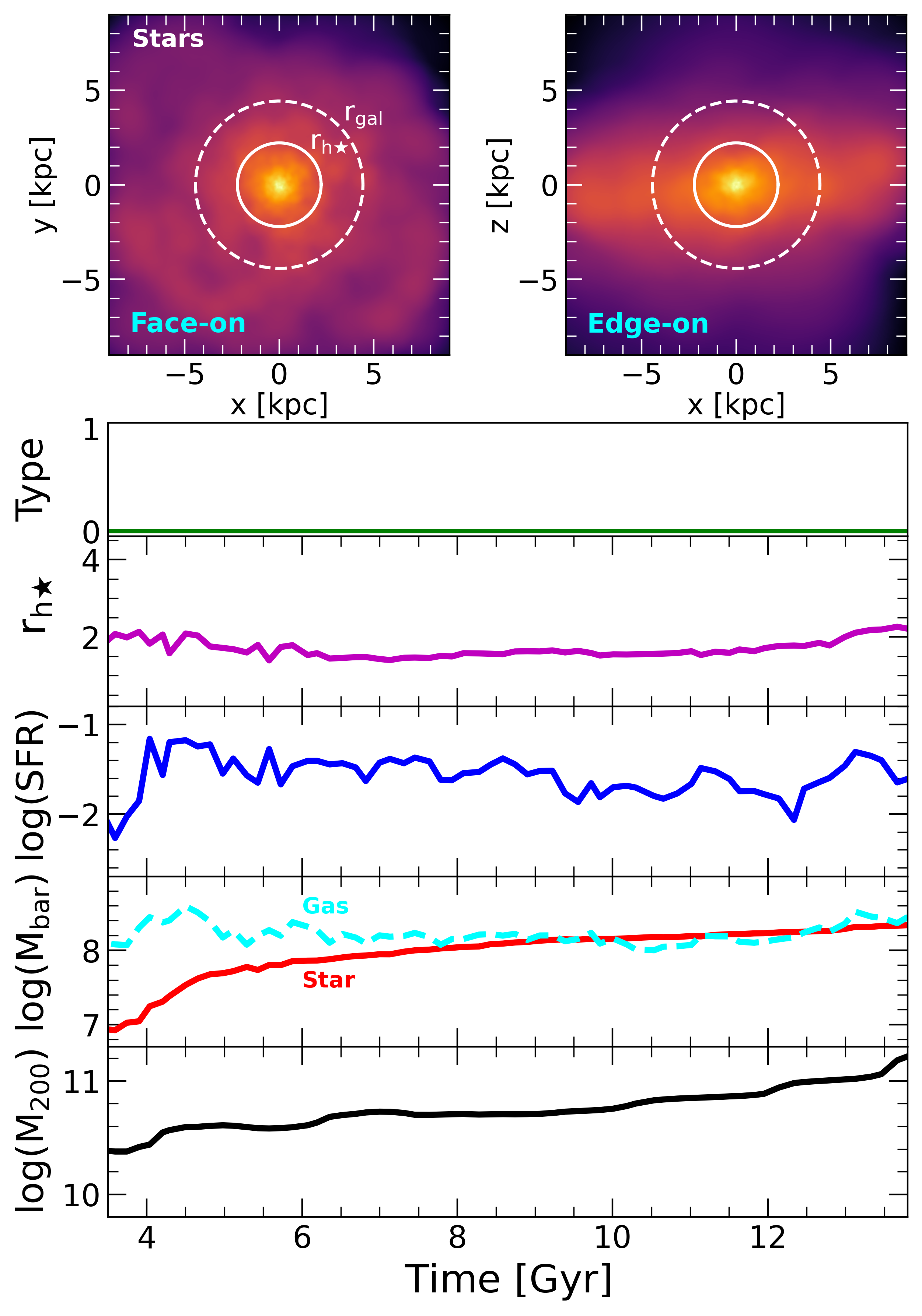}
\includegraphics[width=0.325\textwidth]{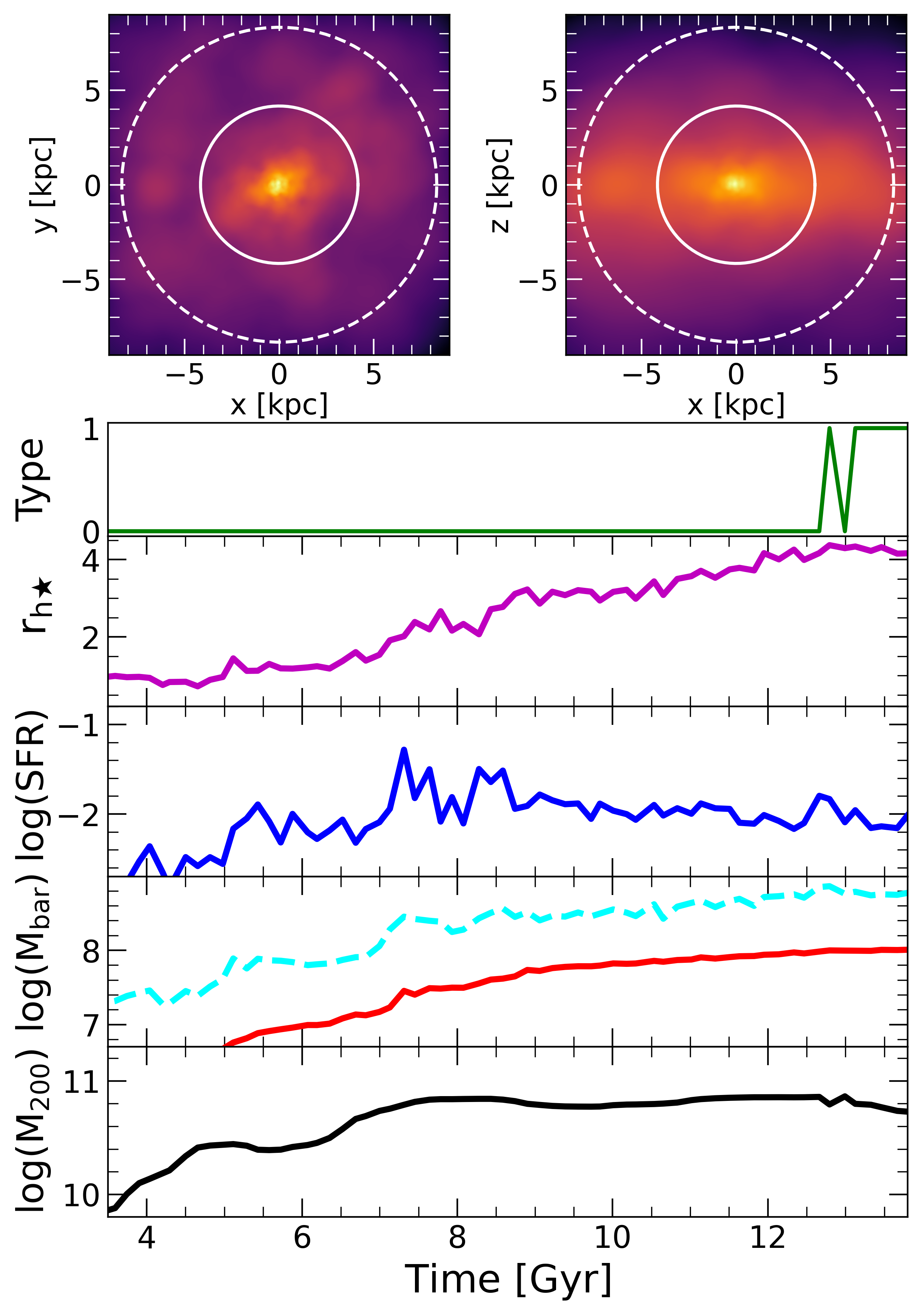}
\includegraphics[width=0.325\textwidth]{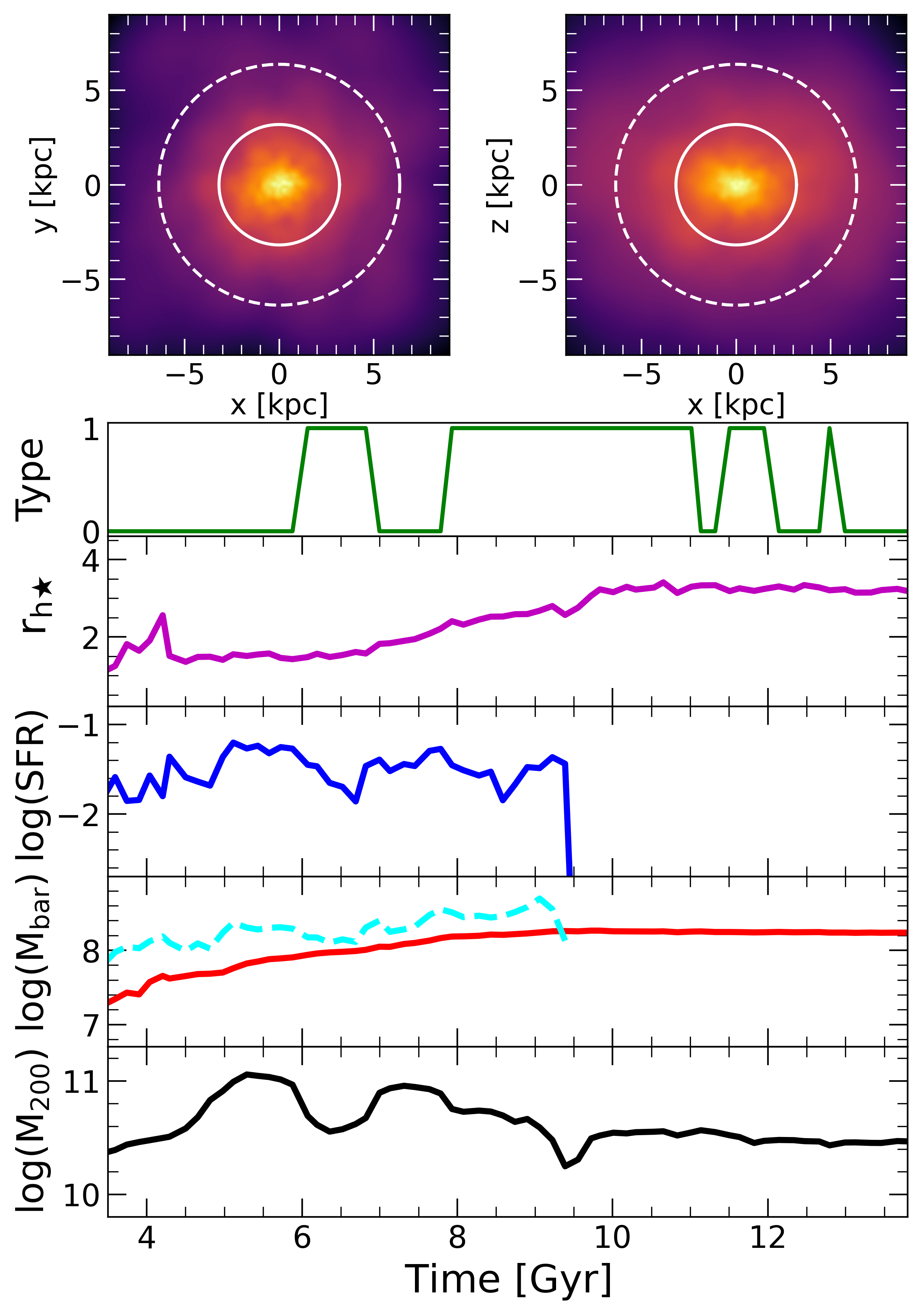}
\caption{Examples of inefficient simulated UDGs, highlighted with green unfilled squares in Fig.~\ref{fig:mhalo_mstr}. Left and center panels are star-forming and the galaxy on the right panel is quenched. In the top panels, face-on and edge-on views of each galaxy are shown and in the bottom the evolution of some properties: type (central:0, satellite:1), 3D stellar half-mass radius ($\rm{r_{h,\star}}$/kpc), star-formation rate ($\rm{SFR}$/$\rm{M_{\odot} \, yr^{-1}}$), baryon mass within $r_{\rm gal}$ (stars in red and gas in cyan, in $\rm{M_{\odot}}$) and halo mass ($\rm{M_{200}/M_{\odot}}$) (dark matter in black and $\rm{M_{200}}$ in grey).}
\label{fig:examples}
\end{figure*}

The right panel of Fig.~\ref{fig:mhalo_mstr} shows a similar trend for our most massive quenched UDGs to fall below the median $\rm{M_\star - M_{200}}$ relation of non-UDGs, at least for the more massive objects with $\rm{M_{\star} \geq 2 \times 10^8 ~ M_{\odot} }$. For our less massive quenched UDGs the trend disappears and UDGs merge to the general stellar mass - halo mass relation. The interpretation of the trends for quenched galaxies is more complex than for star-forming objects, in part, due to tidal evolution. Since the large majority of systems in the quenched subsample are satellites, additional scatter is expected since tidal stripping can change $\rm{M_\star}$ without affecting $\rm{M_{200}}$ in our pre-infall definition of halo mass. In addition, $z_{50}$ is now clearly affected by infall time and it is not longer only characterizing the assembly of each individual halo. Nevertheless, most of the massive quenched UDGs end up forming less stars given their halo mass by factors $\geq 2$, in agreement with findings for the star-forming sample. As expected also from the left panel, quenched UDGs that scatter to the right of the $\rm{M_\star - M_{200}}$ relation have typically later halo assembly times (color map).\\ 

The trend between halo formation time and an inefficient stellar build up is confirmed more clearly in Fig.~\ref{fig:delta_m*}, where for all star-forming (left panel) and quenched (right panel) galaxies, we show the correlation between the formation redshift $z_{50}$ and $\Delta_\star$, corresponding to the vertical departure from the best fit stellar-mass halo-mass relation. UDGs (blue stars / red circles) and non-UDGs (gray) symbols follow a similar trend, where later assembled halos show the largest departure downwards of the relation (large positive $\Delta_\star$). The relative bias in formation time between UDGs and non-UDGs is also illustrated on the top histograms showing the $z_{50}$ distribution. On the other hand, while we find systematic trends on the formation time of the halo for UDGs and non-UDG objects, we find that the stellar assembly (quantified through the time/redshift of formation of half the stellar mass, $t_{50,\star}$ or $z_{50,\star}$) presents a similar distribution for UDGs and non-UDGs.\\

To better understand what is happening, Fig.~\ref{fig:examples} shows 3 examples of inefficient UDGs, two star-forming and one quenched, which have all been highlighted by a large square symbol in Fig.~\ref{fig:mhalo_mstr}. Both star-forming cases are gas-rich and with a stellar component that builds up rather slowly, but the rest of their formation/evolution has been quite different. The UDG on the left panel is a typical case in our sample, inhabiting a central halo with large spin $\lambda \sim 0.06$ and a halo mass $\rm{M_{200} \sim 10^{11} ~ M_\odot}$. Note the steady build-up of dark matter since $t \sim 9$ Gyr (bottom panel), which seems steeper than the stellar mass growth (red curve in the fourth row). \\

An interesting case is shown in the middle panels of Fig.~\ref{fig:examples}. This UDG starts a substantial increase in stellar half-mass radius at $t \sim 7$ Gyr without changes in its halo mass. We have traced this back to a loose dwarf-dwarf interaction that lends additional gas to this galaxy, which cools and is added to a very extended gaseous disk (expanding until a radius $\sim 25$ kpc). In a way, this formation is reminiscent of the Malin-1 analogs reported in the Illustris simulation \citep{Zhu2023}. Notice that at $z=0$ this galaxy lies close to the interacting dwarf and is flagged as the ``satellite" of the interacting dwarf (bottom panel shows that the virial mass of the host system is also $\rm{M_{200} \sim 10^{11} ~ M_\odot}$ confirming the dwarf-dwarf scenario).\\

The quenched UDG example in the right panel of Fig.~\ref{fig:examples} represents a typical case where the progenitor of the UDG infalls at $t \sim 5.3$ Gyr onto a group-mass system and gas is removed rather suddenly at $t \sim 9.5$ Gyr. Star formation is quenched abruptly and the stars seem to expand slightly after quenching (second from top panel), presumably due to tidal heating and changes in the potential due to removal of the gas, leaving a quenched extended dwarf galaxy by $z=0$. Note that this galaxy is gas-dominated at infall, meaning that after the gas is removed by environmental effects, a low $\rm{M_\star}$ is left panel compared to its large dark matter content at infall. This case is representative of most quenched UDGs scattering downwards the $\rm{ M_\star - M_{200}}$ relation.\\

Two factors make a comparison of our predictions with observations difficult. First is the uneven availability of observational constraints for the dark matter content of UDGs, which include only a few observed UDGs with kinematical information of GCs or stars, heavily biased in the selection (for example, selecting those with the largest numbers of GCs or within a given environment). The second factor is that kinematics only constrain the mass within the optical extent of the dwarfs, and strong extrapolations are needed to infer halo mass \citep[e.g., see discussion in ][]{Kravtsov2024}. Taking those factors into account and as discussed in Sec.~\ref{sec:intro}, several observations suggest that UDGs, or a fraction of them, may inhabit overly-massive dark matter halos, in agreement with our results.\\

For example, the predictions from TNG50 indicates that the amount of star formation suppression for the more massive UDGs results in typically factors $\sim 2 \rm - 5$ times less $\rm{M_\star}$ at a fixed halo mass, with some more rare cases with even a larger inefficiency. This seems in rough agreement with the deviations inferred in \citet{Zaritsky2023} for the observed UDG population, although their typical inefficiency may be larger than in our sample. Interestingly, for lower mass UDGs with $\rm{M_\star < 10^8 ~ M_{\odot}}$ the \citet{Zaritsky2023} sample indicates no difference in the $\rm{ M_\star - M_{200}}$ relation, in good agreement with our results. Overall, considering the lack of spectroscopy and uncertainties involved in halo mass extrapolation in these observations, the agreement of our predictions with the general population of UDGs is highly encouraging.\\ 

Do we find simulated objects as extremely inefficient as observed? This question is more difficult to answer. UDGs with stellar mass suppression beyond factors $\sim 5$ are quite rare in TNG50 but they seem more common in \citet{Zaritsky2023}. Similarly, several of the Virgo UDGs with known kinematics for their GCs system suggest dark matter masses consistent with more massive MW-like halos \citep{Toloba2023}, which we were unable to reproduce in TNG50 even when making mock observations of our simulated systems that mimic the exact same observational techniques \citep{Doppel2024}. \\

One might argue that uncertainties and biases in observational samples might be enough to reconcile the results. However, if the frequency of strongly inefficient UDGs (objects with $\sim 10$ times less stars than expected for their $\rm{M_{200}}$) is confirmed with other samples and in similar environments as those included in TNG50, it may suggest that there are additional mechanisms at play that are not well reflected in this simulation (or any other simulation reported so far). For now, we have showcased some of our most extreme examples (see Fig.~\ref{fig:examples}) in an attempt to provide insights as to how the most inefficient UDGs could have formed. \\
 

\begin{figure*}
\centering
\includegraphics[width=\textwidth]{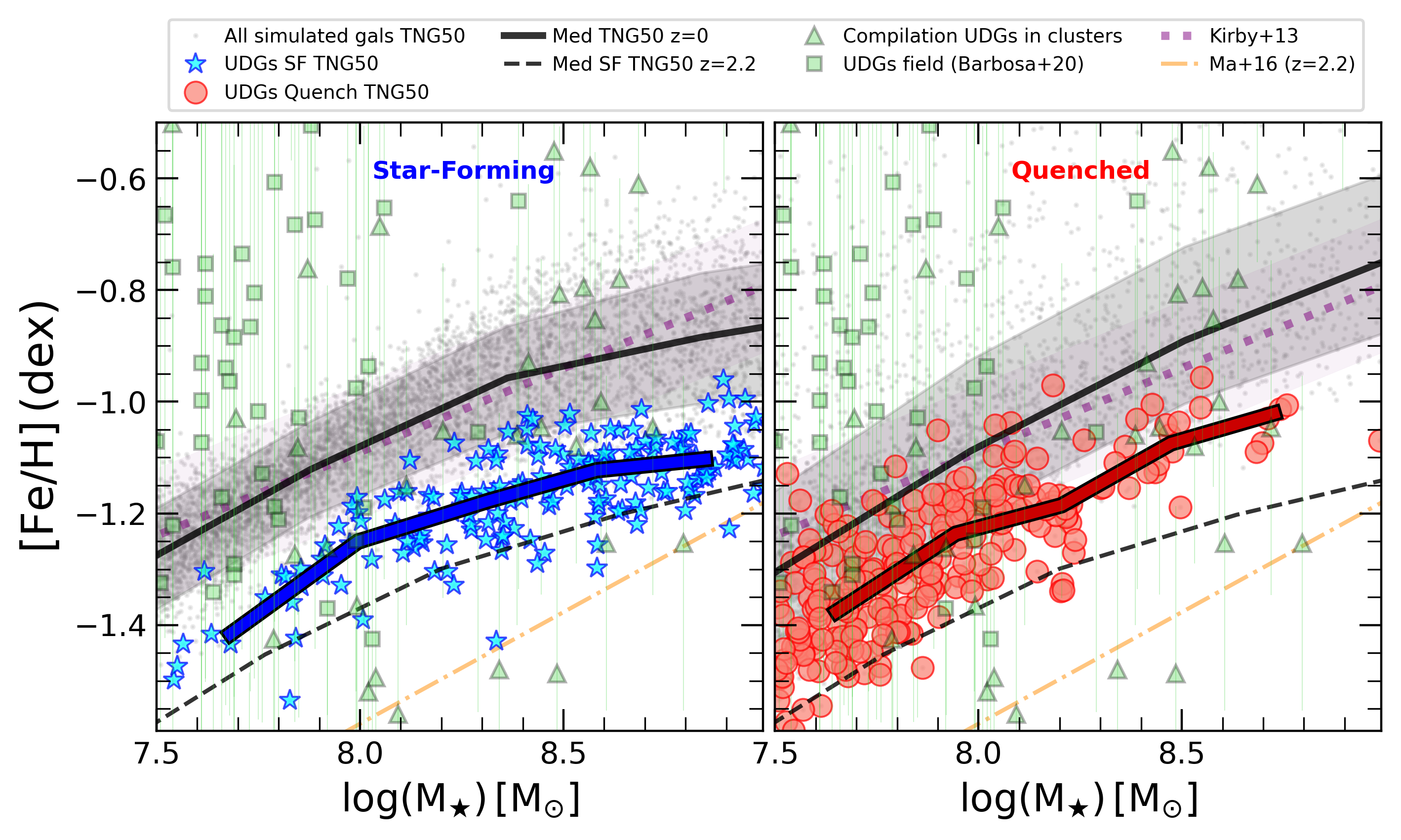} 
\caption{Relation between stellar mass and metallicity ([Fe/H]) for all simulated galaxies in the TNG50 simulation (grey dots), separated by star-forming (left panel) and quenched (right panel). The median relation for each population is highlighted with the thick black solid line in each panel. The sample of simulated UDGs from TNG50, star-forming and quenched, are indicated with blue stars and red circles in each panel, with their respective median profile shown by thick solid lines. A compilation of different observational data of UDGs is highlighted with green symbols \citep{Smith2009, FerreMateu2018, Gu2018, Barbosa2020, FerreMateu2023}. The mass-metallicity relation from \citet{Kirby2013} is indicated in the dotted purple lines (thick dashed, dotted, and thin dashed, respectively). For reference we also include the simulated mass-metallicity relation at $z=2.2$ (dashed black line) and the observed relation at $z=2.2$ from \citet{Ma2016} (orange dot-dashed). Simulated UDGs are typically more metal-poor than non-UDG galaxies of similar stellar mass and show less overall metallicity dispersion than the observed samples.}
\label{fig:mass_FeH}
\end{figure*}

\section{Metallicities and metallicity profiles in simulated UDGs}
\label{sec:metals}

Besides the dark matter halos they inhabit, recent observations have been able to provide new constraints on the stellar population of UDGs that may be compared against our theoretical predictions. We start by studying the stellar mass - metallicity relation in Fig.~\ref{fig:mass_FeH}, where metallicity is here quantified by the iron-to-hydrogen fraction [Fe/H] calculated from all stellar particles within the galaxy radius $\rm{r_{gal} = 2 r_{h,\star}}$. Note that the simulated metallicities need to be re-normalized in order to fit the median mass-metallicity observed. Here we use \cite{Kirby2013} to describe the observed relation (see purple dotted line). We apply a -0.52 dex shift to all simulated galaxies, which is calculated such that the median metallicity of TNG50 at $\rm{M_\star \sim 10^{8} ~ M_{\odot}}$ coincides with \citet{Kirby2013} using all galaxies (star-forming + quenched).\\

As before, in Fig.~\ref{fig:mass_FeH} we divide our sample into star-forming (left panel) and quenched (right panel) samples. Non-UDG galaxies in the simulation box are shown with gray dots and UDGs are highlighted with blue stars or red circles for star-forming and quenched, respectively. The first remark from this figure is that simulated UDGs are systematically more metal-poor at a given $\rm{M_\star}$ compared to non-UDG galaxies. This is shown more clearly by the comparison between solid blue (red) lines indicating the median of star-forming (quenched) UDGs and the solid black line in each panel, calculating from all non-UDG galaxies in each subsample. In general, the simulation predicts a clear gradient towards lower metallicity at fixed $\rm{M_\star}$ as the half-mass radius increases, making the $\sim 0.2$-dex lower metallicity in our UDG sample a consequence of this continuous trend (see Fig.~\ref{fig:app_MZR_size} in the Appendix). This agrees with the lower metallicities associated in general with low-surface brightness galaxies in TNG50 reported recently in \citet{Tang2024}, a trend that might arise as a result of late gas accretion in the most extended galaxies, resulting on stars formed from less polluted gas \citep{SanchezAlmeida2018}. \\

The second important point to notice in Fig.~\ref{fig:mass_FeH} is that the scatter in simulated UDGs is substantially smaller than that reported in observational UDG studies (green symbols). While the population including the non-UDGs (gray dots) manages to cover the range spanned by the green symbols, those classified as UDGs in our simulated sample show a more limited [Fe/H] range. We have confirmed that simulated non-UDGs that scatter above the r.m.s in the mass-metallicity relation correspond to dwarfs with substantial tidal disruption (having lost half or more of their stellar mass). However, none of those objects are sufficiently extended to be classified as UDGs in our sample.\\

Despite the reduced scatter in our simulations, the systematic downwards shift of our UDG sample compared to the non-UDGs causes several of our objects to scatter quite far down the mass-metallicity relation, some reaching values that are consistent with the average simulated mass-metallicity relation at $z=2.2$ (orange dash-dotted line)\footnote{Note that we have applied the same $-0.4$-dex shift to the $z=2.2$ relation used to normalize the $z=0$ metallicities. However, the simulated median mass-metallicity relation at $z=2.2$ is still higher than the observed relation, as reported by \citet{Ma2016}}. We find that these very metal-poor UDGs can be both, star-forming and quenched, meaning that there is not a direct link between an early halt of star-formation and their low metallicity, as seems to be the case for some observed UDGs \citep[see e.g., ][]{FerreMateu2023,Buzzo2024}. Moreover, we find no evidence for these low metallicity UDGs to be particularly more inefficient at forming stars (larger $\Delta_\star$) than the comparatively more metal-rich UDGs. We conclude that for the formation scenario of UDGs in TNG50, low metallicities are not an indication of an excess dark matter halo compared to the stellar component but instead seem simply the result of a trend predicted in the simulation where at fixed stellar mass, more extended galaxies have lower metal content (see Fig.~\ref{fig:app_MZR_size} in the appendix).\\

\begin{figure}
\centering
\includegraphics[width=\columnwidth]{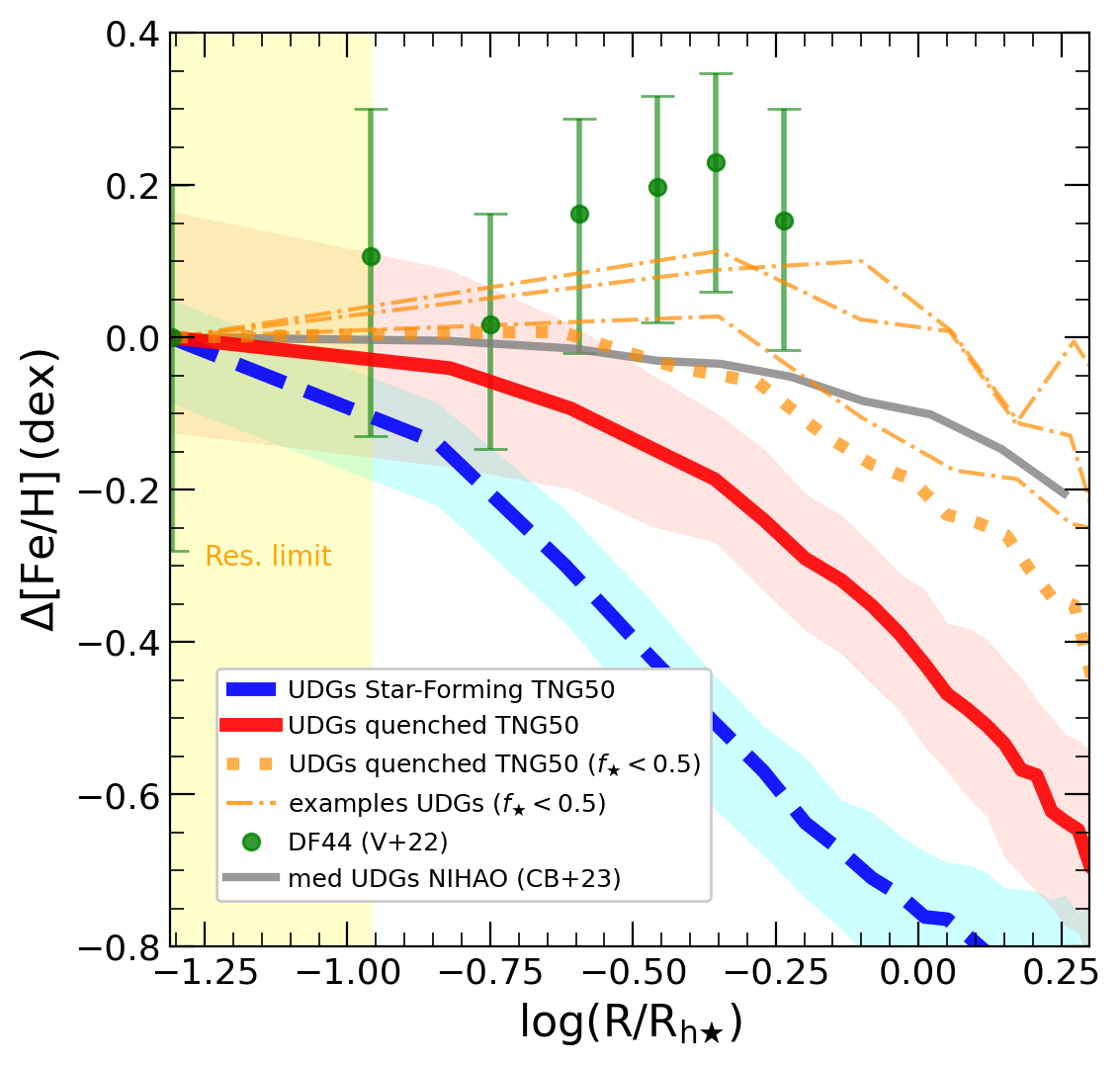}
\caption{Median metallicity profiles for the sample of simulated star-forming (blue) and quenched (red) UDGs in TNG50. The shaded regions correspond to the 25th and 75th percentiles in each distribution. We predict metallicity gradients that are steeper than the scenario where UDGs are formed from strong outflows: gray thick curve shows the median metallicity profile of UDGs from the NIHAO simulations  \citep{CardonaBarrero2023}. Simulated UDGs with tidal stripping also show flatter profiles in TNG50, with the orange dotted line indicating the median of all UDGs that have lost at least $50\%$ of their stars. Three examples of these most extreme stripped cases are shown individually with thin dot-dashed lines, which all seem in better agreement with the NIHAO results as well as with the observational data in DF44 by \citet{Villaume2022} shown with green symbols. This suggests metallicity profiles as promising candidates to help identify the formation scenario of UDGs.}
\label{fig:metal_profile}
\end{figure}

\subsection{Metallicity profiles}
\label{ssec:metal_profiles}

Fig.~\ref{fig:metal_profile} shows the median projected metallicity ([$F_e/H$]) profile of UDGs as a function of projected galactocentric distance R. To ease the comparison of different galaxies, we normalize the vertical axis to the value at the center of each object, leading to the relative metallicity distribution with respect to the central value. The horizontal axis is also normalized by the projected stellar half-mass radius $\rm{R_{h,\star}}$ of each UDG. The median of our simulated UDG sample is shown by the dashed blue and solid red lines for star-forming and quenched subsamples, respectively, and the shaded area indicates the 25th and 75th percentiles of both distributions. The simulation predicts a well-defined metallicity gradient for both kinds of UDGs, with metallicities near the edge ($\rm{\sim 2 R_{h,\star}}$) that are typically $\sim 0.8$ - $1$ dex lower than at their centers. The gradient is slightly steeper for the star-forming UDGs. We have checked (not shown) that both subsamples follow well the trend predicted for star-forming and quenched non-UDGs in TNG50, indicating that only the overall metal content but not the metallicity distributions within galaxies seem to change with stellar size in the simulation.\\

Interestingly, simulations where UDGs form by intense outflows driven by stellar feedback predict, instead, a flatter profile. This is shown by the grey solid line in Fig.~\ref{fig:metal_profile} highlighting the median metallicity profiles reported by \cite{CardonaBarrero2023} in the NIHAO simulations, where the diffuse nature of UDGs is better explained by stellar feedback effects \citep{DiCintio2017} and not an excess of spin, like in our sample. Shallower profiles in this case make sense, as the same gas flows that are turning these dwarfs diffuse may also cause stellar mixing and re-distribution of metallicity profiles making them shallow \citep{Mercado2021}. This suggests that metallicity profiles may be a way to disentangle the formation path of UDGs.\\

Some caveats may apply. First, while the metallicity trends are different {\it on average} for different formation paths, individually, some UDGs might not conform with the rest of the sample. For example, the individual metallicity profiles shown in the \citet{CardonaBarrero2023} sample have a few of the NIHAO UDGs nicely tracking the median trend in the TNG50 sample. The origin for their UDGs with more steep metallicity profile is unclear, but they would be perhaps mistakenly identified with a more gentle formation scenario as in TNG50 if the trend is applied directly to individual cases. Similarly, tidal stripping can also introduce changes to the metallicity distribution. The orange thick dotted curve shows the median metallicity profile for UDGs in TNG50 that show substantial stellar mass loss due to tidal stripping (the fraction of retained stellar mass is less than half $f_\star < 0.5$), and shows a flatter profile with radius that comes closer to the NIHAO average predictions. In fact, we also show the 3 examples of these most extreme tidal disruption cases (thin dot-dashed orange curves), with a stellar mass retention $f_\star \sim 0.4$, which may be indistinguishable from several of the NIHAO UDGs.\\

The second caveat to bear in mind is the difficulty of measuring metallicity profiles in observations for objects with such low surface brightness. So far, only one measurement has been reported for DF44 \citep{Villaume2022}, and it suggests a rather flat or even increasing profile in the inner regions. Such a trend would suggest that it is unlikely that DF44 is well represented by simulated UDGs from the TNG50 sample, although some tidal disruption could help reconcile the trend (light blue dot-dashed curves). It is possible that more observational measurements of the metallicity distributions in the UDGs could shed some light on their formation mechanism, however in order to assemble a large enough sample with $\sim$dozen(s) may require a serious commitment from the community, as each individual UDG may take $\sim 1$ full night of observations with the best available telescopes. Of note is also the fact that spectroscopic surveys necessarily favor the studies of the slightly brighter (or higher surface brightness) objects even within the UDG category with the risk of introducing unwanted biases in the interpretation of average results \citep[see discussion in][]{Gannon2024}.\\

\begin{figure}
\centering
\includegraphics[width=\columnwidth]{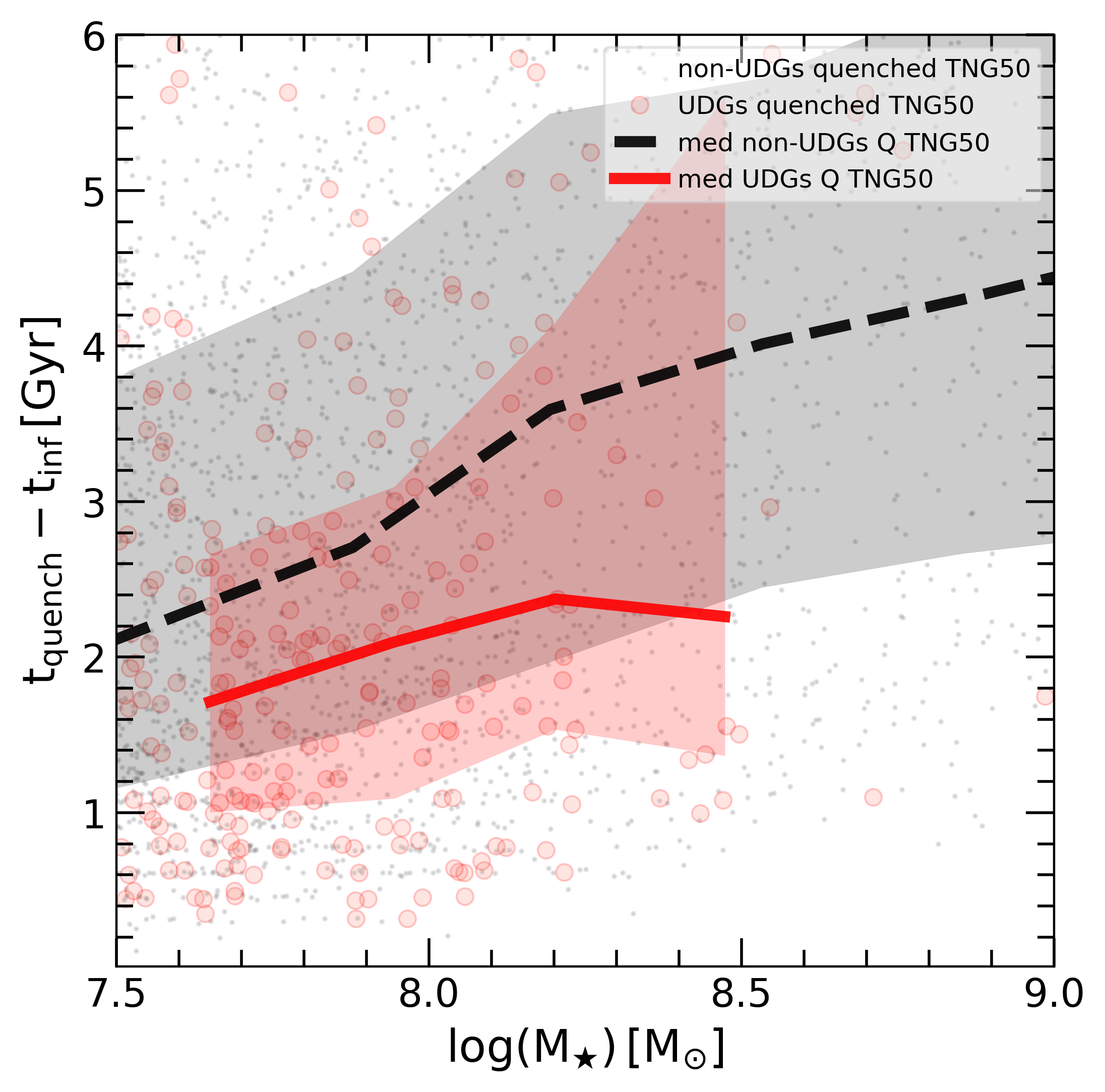}
\caption{Quenching time-scale as a function of stellar mass for quenched non-UDG dwarfs (gray) and quenched UDGs (red) in TNG50. The medians of each population are indicated with thick lines, while the shaded regions are the 25th to 75th percentiles. On average, the UDGs (red line) have shorter quenching timescales than non-UDGs (dashed black line) at a given stellar mass.} 
\label{fig:quech_timescales}
\end{figure}

\subsection{Quenching time-scales}
\label{ssec:timescales}

To finalize characterizing the stellar population properties of simulated UDGs, we study in Fig.~\ref{fig:quech_timescales} the typical timescales over which the quenching of star formation occurs in the TNG50 simulations. The quenching time, $t_{\rm quench}$, is defined as the time where the specific star formation rate of a given simulated galaxy falls below $\rm{sSFR = 10^{-11} ~ yr^{-1}}$ \citep{Wetzel2012}. We then show the difference between $t_{\rm quench}$ and the infall time, where infall time $t_{\rm inf}$ is defined as the first time a galaxy becomes a satellite of another host halo. Note that in the case of field UDGs that are quenched, they can always be attributed to backsplash orbits \citep{Benavides2021}, for which we may still define the infall time as the time they interacted with their host system. Red symbols correspond to quenched UDGs which are compared against non-UDGs in the same mass range (black symbols).\\

\begin{figure}
\centering
\includegraphics[width=\columnwidth]{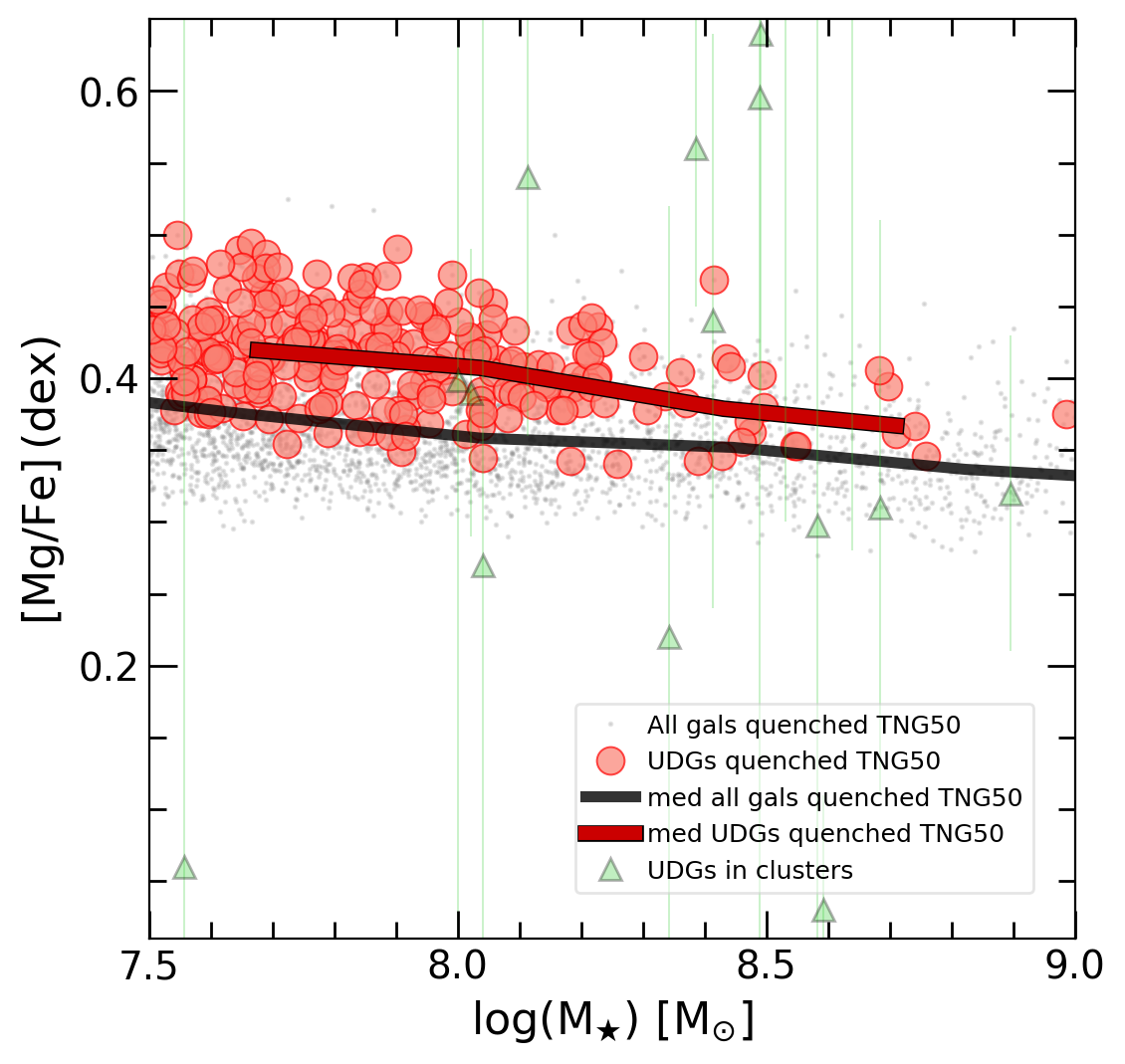}
\caption{Magnesium to Iron ([Mg/Fe]) abundance as a function of stellar mass for simulated non-UDG dwarfs (gray) and quenched UDGs (red symbols). The median relation for each sample is highlighted with the thick solid black line and red line, respectively. UDGs are predicted to have an elevated [Mg/Fe] ratio compared to non-UDGs of the same mass. A sample of observational UDGs by \citet{FerreMateu2018, FerreMateu2023,RuizLara2018} are included with green symbols}
\label{fig:MgFe_Mstar}
\end{figure}

We find that although the distribution of infall times is very similar for both populations (not shown), the distribution of the quenching timescales is different, being on average shorter for UDGs than non-UDGs. This can be seen by the median trends highlighted by the red solid and black dashed lines for UDGs and non-UDGs, respectively as well as by the integrated histogram in the vertical axis. At all masses the medians differ by $\sim 1.3$ Gyr, with UDGs being systematically below the non-UDG sample. In addition, we see a slight relation with stellar mass, tending to larger quench timescales for more massive galaxies \citep[in agreement with ][]{Mistani2016} and perhaps indicating the more robust gravitational potential of the more massive dwarfs to counteract environmental effects.\\

We show in the Appendix~\ref{ssec:ram_press_efect} that a simple ram-pressure stripping model would be able to explain a faster gas exhaustion in UDGs compared to non-UDG objects. In this model, we assume a dwarf galaxy with $\rm{M_\star = 10^{8.2} ~ M_{\odot}}$ and use the average dark matter halo mass for UDGs and non-UDGs at this $\rm{M_\star}$ assuming for both a concentration $c=15$. We then also consider a disk made of gas using the average gas content for UDGs and non-UDG galaxies. We find that, for galaxy cluster environments, the ram-pressure force is able to totally strip gas from UDGs while for non-UDG galaxies, the larger baryonic concentration helps retain gas within $\rm{R_{h \star}}$. Similarly, for group environments, gas would be stripped outside of $\rm{R_{h \star}}$ for UDGs, while for non-UDGs gas may be retained out to $2 - 3 \rm{R_{h \star}}$. It is then possible that the shorter timescales for quenching can be explained as a result of a decreased gravitational restoring force in UDGs compared to typical ram-pressure forces within groups and clusters.\\

Consistent with the shorter timescales for quenching, we show in Fig.~\ref{fig:MgFe_Mstar} that simulated quenched UDGs in TNG50 are systematically more alpha-enriched relative to other quenched non-UDG objects with similar mass. Compared to the few observational estimates of [Mg/Fe] available in the literature \citep{FerreMateu2018, FerreMateu2023, RuizLara2018}, it looks that once again simulated objects (UDGs and non-UDGs) exhibit insufficient scatter in the abundance of alpha-elements. Yet, the trend found with simulated UDGs having systematically larger [Mg/Fe] at fixed $\rm{M_{\star}}$ should be robust to shortcomings in the metallicity modeling in TNG50.\\

\section{Summary}
\label{sec:concl}

We use the TNG50 hydrodynamical cosmological simulation to study the predictions for UDGs in the stellar mass range $\rm{M_\star = 10^{7.5} - 10^{9}~ M_{\odot}}$, with specific emphasis on the dark matter content and stellar population properties. In this work, we analyze the dark matter halos, metallicity, $\alpha$/Fe element ratio, metallicity profiles, and quenching timescales for a sample of simulated UDGs divided in two subsamples: star-forming and quenched. We also use control samples from TNG50 with galaxies of the same stellar mass but consider ``normal" in their sizes (non-UDGs). Our main results can be summarized as follows.\\

UDGs are inefficient at forming stars, in particular the star-forming subsample. At a given virial halo mass, UDGs typically form $2$-$5$ times fewer stars compared to the overal non-UDG population, with some outliers showing suppression factors as large as $\sim 10$. For quenched UDGs the trend is less systematic compared to non-UDGs, but still several quenched UDGs show suppression factors $5$-$10$ with respect to the stellar mass expected for their dark matter halo, in particular for $\rm{M_\star \geq 10^{8.2} ~ M_{\odot}}$. We trace the lower stellar content at a given halo mass in our UDG sample to later formation times for their halos. Our most stellar-depleted objects form in halos that are assembled more recently, and therefore the stellar content has not had enough time to catch up to the increased $\rm{M_{200}}$. In general, our results agree with the idea that UDGs are inefficient galaxies, as suggested by the SMUDGes survey \citep{Zaritsky2022}, although we find smaller suppression factors. No difference in the stellar mass - halo mass relation is predicted in our simulated UDGs below $\rm{M_\star \sim 10^8 ~ M_{\odot}}$ also in good agreement with observations \citep{Zaritsky2023}.\\

The stellar assembly of UDGs in TNG50 seems in agreement with recent observational constraints for UDGs in low and high-density regions \citep{FerreMateu2023}. Interestingly, for the quenched population, we find that the typical quenching times in UDGs is shorter on average by $1.3$ Gyr than in non-UDGs of the same stellar mass. Increased ram-pressure stripping due to their more diffuse nature is identified as a possible culprit of this difference.\\

In TNG50, UDGs are on average metal-poor, holding at fixed $\rm{M_\star}$ (on average) $60\%$ less metals ([Fe/H]) than simulated non-UDGs of the same stellar mass. However, we notice that simulation predictions (for UDGs and non-UDGs) fail to represent the appropriate level of scatter in [Fe/H] seen in observations. Because of the general trend towards low metallicities at fixed $\rm{M_\star}$, some UDGs resemble the mass-metallicity relation of the simulation at $z \sim 2$. This trend is consistent with the finding reported for some extreme metal-poor UDGs \citep{FerreMateu2023, Buzzo2024}, which are interpreted as prime candidates to be ``failed galaxies''. However, in our simulation their low metallicity seems simply the reflection of the scatter around an already low value and does not highlight or pick-up extreme objects in terms of their assembly history.\\

We identify a new indicator able to distinguish certain aspects of the formation mechanism of UDGs: metallicity profiles. For the TNG50 simulations where the large extended sizes of UDG analogs are explained as a combination of high-spin halos and a slight excess of halo virial mass \citep{Benavides2023}, we find a well-defined declining metallicity profile with distance from the center of the galaxy. This is seen in both our samples, the star-forming and quenched UDGs, with an annulus-averaged metallicity profile that may decrease $\sim 0.6$-$0.8$ dex from the center to twice their half-mass radius. This is in contrast with the predictions of more shallow metallicity profiles from the simulations where UDGs formed due to intense feedback outflows as for example, in the NIHAO simulations \citep{CardonaBarrero2020}. Tidal stripping in the TNG50 simulated UDGs seems to induce also flattening of the profiles, but such objects would have elevated metallicities at a given $\rm{M_\star}$ offering the possibility to combine metallicity profiles with total metallicity to shed some light on the formation path of observed UDGs.\\

Observationally, metallicity profiles are taxing to measure, especially for the most extreme UDGs with the lowest surface brightnesses, but at least one object has been reported in the literature: DF44, which seems to indicate a quite flat [Fe/H] profile, at least in the inner regions. In the future, the availability of more data like in DF44 may offer the possibility to learn some aspects of the formation of these dwarfs and their relation to the rest of the non-UDG dwarf population.\\

\section*{Acknowledgments}
The authors would like to thank Annalisa Pillepich and the whole TNG50 team for early access to the simulation data. The authors are also thankful to Dennis Zaritsky, Eric Peng, Duncan Forbes, and 
Aaron Romanowsky for helpful discussions that helped shape this paper. We would also like to thank Anna Ferr\'e-Mateu and Salvador Cardona-Barrero for sharing data related to the mass assembly history and metallicity profiles of their samples. JAB and LVS is grateful for partial financial support from NSF-CAREER-1945310 and NSF-AST-2107993 grants. MGA acknowledge financial support from CONICET through PIP 11220170100527CO grant. Computations were performed using the computer clusters and data storage resources of the HPCC, which were funded by grants from NSF (MRI-2215705, MRI-1429826) and NIH (1S10OD016290-01A1).

\section*{Data Availability}

This paper is based on halo catalogs and merger trees from the Illustris-TNG Project \citep{Nelson2019TNG, Nelson2019TNG50}. These data are publicly available at \href{https://www.tng-project.org/}{https://www.tng-project.org/}. The main properties of the UDG and non-UDG dwarf galaxy samples, and other products included in this analysis, may be shared upon request to the corresponding author if no further conflict exists with ongoing projects. 


\bibliographystyle{aasjournal}
\bibliography{biblio}

\appendix

\section{Dark matter spin halo and scatter on the Mass-Metallicity relation}
\label{app:lambda_and_MZR}

\subsection{Dark matter halos spin ($\lambda$) vs $z_{50}$}
\label{app:lambda_z50}

\begin{figure}
\centering
\includegraphics[width=0.7\columnwidth]{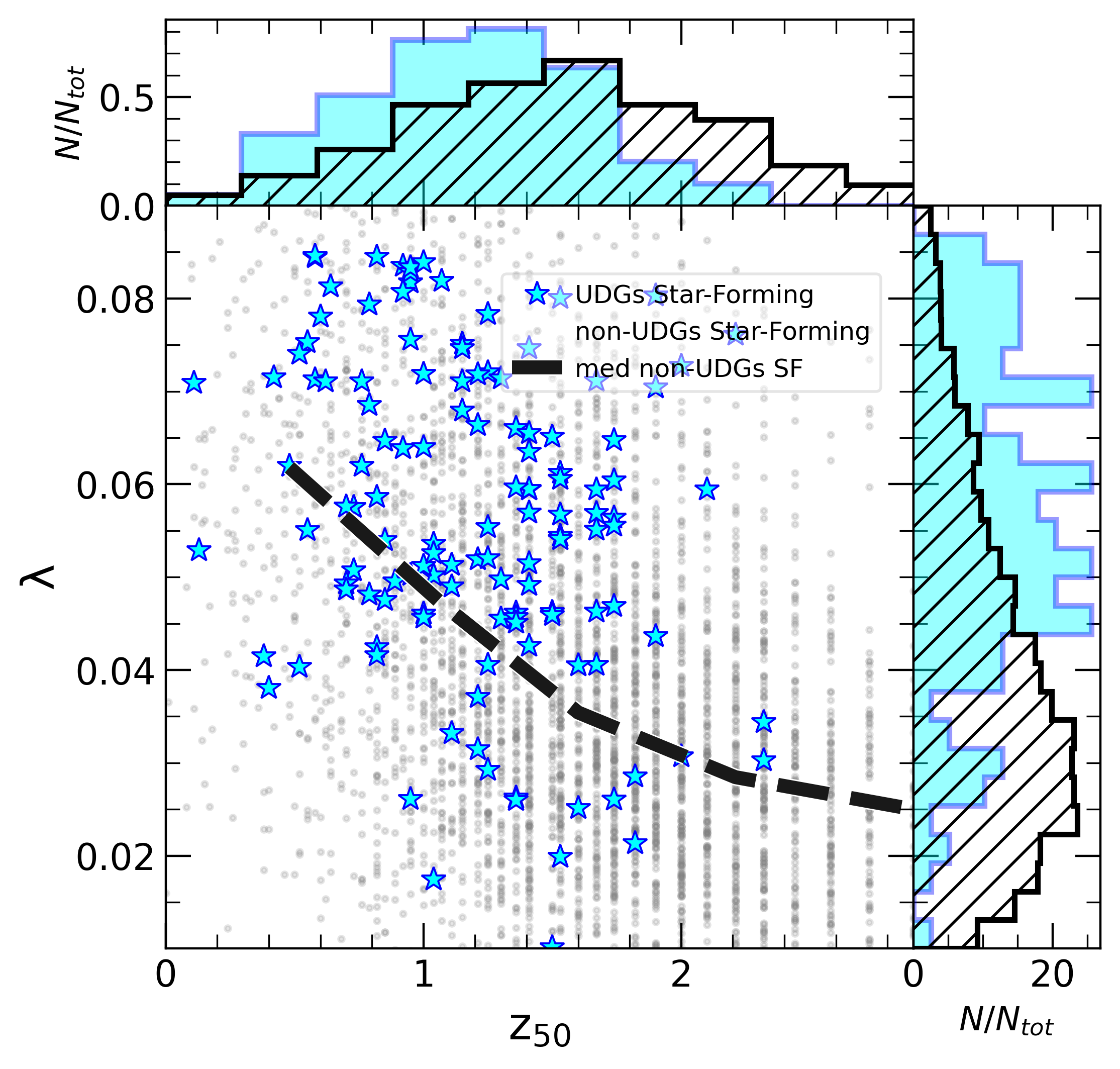}
\caption{Halo spin parameter ($\lambda$) for simulated dwarf galaxies as a function of the dark matter halo formation redshift ($z_{50}$), defined as the redshift at which each halo attains half of their present-day virial mass. Normal star-forming dwarf galaxies (non-UDGs) are shown in grey individual symbols with their median trend indicated by the thick dashed line. Halos that assemble late have typically larger spins. Star-forming UDGs are highlighted in blue star symbols and follow the general trend outlined by the non-UDG dwarfs.}
\label{fig:app_lambda_z50}
\end{figure}

In \citet{Benavides2021, Benavides2023} was shown how simulated UDGs (central and satellites) formed in dark matter halos with a high spin parameter ($\lambda$). Here we include the relationship between this spin parameter and the redshift of dark matter halo formation, $z_{50}$ (see Fig.~\ref{fig:app_lambda_z50}). Dark matter halos that assemble late (lower $z_{50}$) have typically higher spin values. Gray dots indicate all dwarf galaxies in our mass range, with their median relation highlighted in thick dashed black line and blue star symbols highlighting the star-forming UDGs. Note that for simplicity we only include the star-forming sample since they are mostly central halos and their $\lambda$ parameter is less affected by interactions \citep[see ][]{Benavides2023}. 

\subsection{Scatter in the mass-metallicity relation with galaxy size}
\label{app:fstar_colored}

\begin{figure}
\centering
\includegraphics[width=0.7\columnwidth]{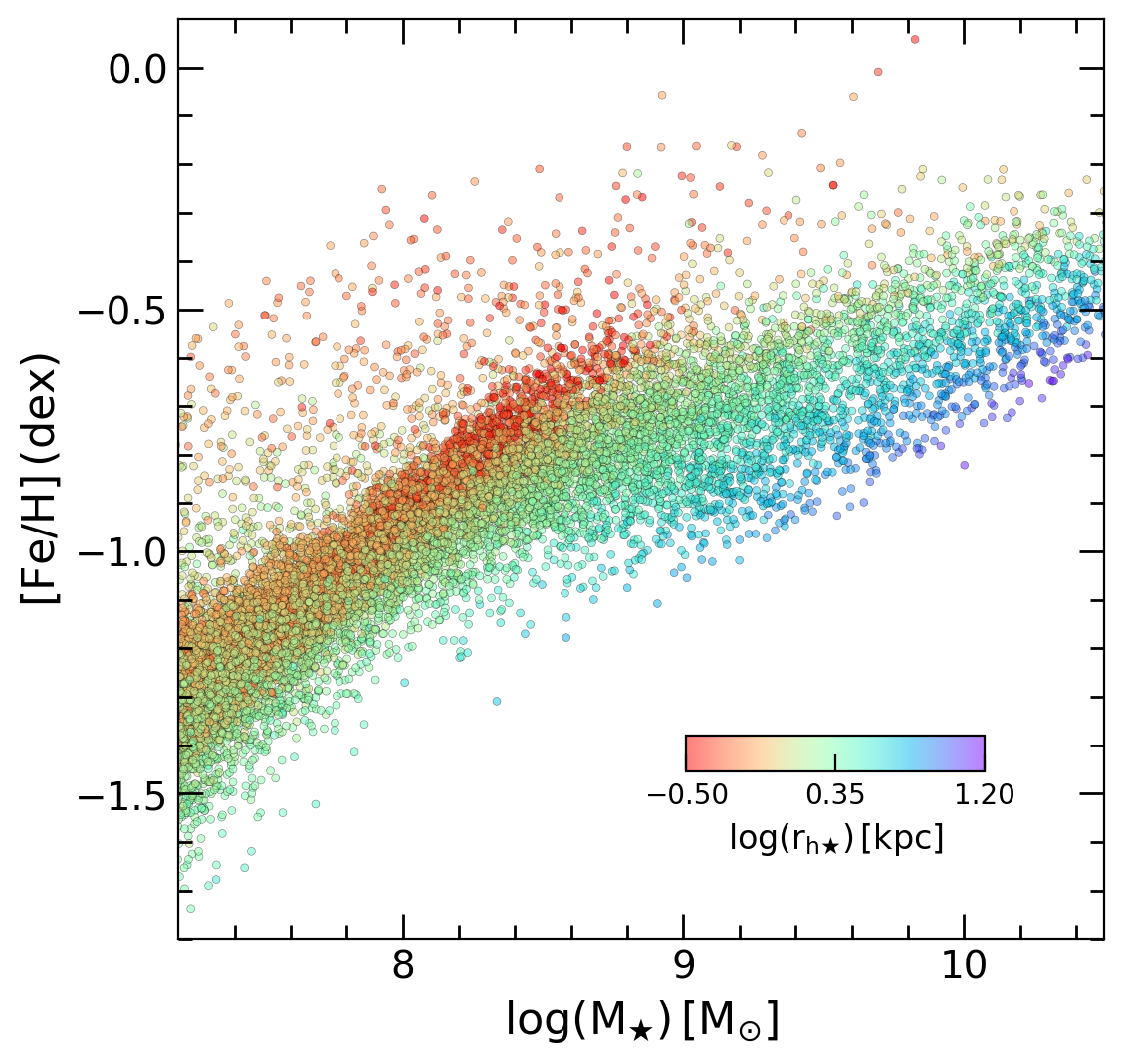}
\caption{Mass-Metallicity relation for all simulated galaxies in TNG50, where the color code indicates the stellar half-mass radius.}
\label{fig:app_MZR_size}
\end{figure}

TNG50 is not able to reproduce the large scatter of UDGs in the mass-metallicity relation, but the simulation does predict a substantial scatter for the population overall. At a fixed stellar mass, Fig.~\ref{fig:app_MZR_size} shows a clear trend with the stellar half-mass radius, indicating that objects with smaller sizes tend to scatter upwards the relation, while galaxies with the most extended sizes scatter downwards. As a result, the metallicities predicted for UDGs are always below the median of the simulation as a whole.\\

\section{How do the size and gas content impact the quenching timescale in UDGs}
\label{ssec:ram_press_efect}

The ram-pressure force acting on a galaxy moving with velocity $v$ in a host media with intra-cluster gas density $\rho_{ \rm ICM}$ is \citep{GunnGott1972}:
\begin{equation}
    P = \rho_{\rm ICM} \, v^2 \ .
\label{eq:rampress}	
\end{equation}
\noindent

This pressure force is counteracted by the self-gravity of the moving satellite galaxy. To calculate the latter, we include a model for dwarf galaxies (UDGs and non-UDGs) inspired by the work of \citet{Abadi1999}. The idea is to compare the ratio of the total vertical force per unit area of the galaxy ($\sigma_{\rm gas} \partial \phi /  \partial z$) against the surrounding pressure ($\rho_{\rm ICM} v^2$), such that if $\sigma_{\rm gas} \partial \phi /  \partial z > \rho_{\rm ICM} v^2$ the galaxy is able to retain the gas. In our case, we have modeled two dwarf galaxies, a UDG and a non-UDG composed of a stellar disk and a gaseous disk within a dark matter halo. For the case of the stellar and gaseous disk, we use an exponential mass profile whose potential, in cylindrical coordinates, is:

\begin{equation}
    \phi_d (R,z) = -2\pi G \Sigma_0 R^2_d \int^{\infty}_0  \frac{J_0(K R) \exp{-k |z|}}{[1+(k R_d)^2]^{3/2}} dk \ , 
\label{eq:disk_pot}	
\end{equation}
\noindent

with $\Sigma_0 = Md/(2\pi R^2_d)$, and $J_0(x)$ the Bessel function of first kind order zero, and the acceleration in the z direction is:

\begin{equation}
    \frac{\partial \phi_d}{\partial z}(R,z) = G M_d \int^{\infty}_0  \frac{J_0(K R) \exp{-k |z|}}{[1+(k R_d)^2]^{3/2}} k dk \ .
\label{eq:accel_disk_z}	
\end{equation}
\noindent

For the dark matter halo, we use an NFW profile \citep{NFW} whose potential could be expressed in the form \citep{Lokas2001}:

\begin{equation}
    \phi_h (r) = -\frac{4 \pi}{3} G g(c) r^3_{\rm vir} \Delta_{\rm vir} \rho^0_c \frac{\ln (1 + r/r_s)}{r} \ , 
\label{eq:nfw_pot}	
\end{equation}
\noindent
with $g(c) = 1 / [ \ln(1+c) - c/(1+c) ]$, where $c$ is the halo concentration, $r_{\rm vir}$ the virial radius, the scale radius $r_s = r_{\rm vir}/c$, $\Delta_{\rm vir}$ the overdensity and $\rho^0_c$ the critical density of the Universe at $z=0$. From this, the vertical force in cylindrical coordinates is:

\begin{equation}
    \frac{\partial \phi_h}{\partial z}(R,z) = K \frac{z}{r} \left( \frac{ \frac{r}{r_s + r} - \ln \left( 1 + \frac{r}{r_s} \right)}{r^2} \right) \ ,
\label{eq:accel_halo_z}	
\end{equation}
\noindent
with $K = \frac{4 \pi}{3} G g(c) r^3_{\rm vir} \Delta_{\rm vir} \rho^0_c$.\\

For the mass density of the disk gas: 
\begin{equation}
    \sigma_{\rm gas} (R) =  \int^{\infty}_{-\infty} \rho_d (R,z) dz = \frac{M_{\rm gas}}{2\pi R^2_{\rm gas}} \exp{-R/R_{\rm gas}}\ .
\label{eq:sigma_gas}	
\end{equation}
\noindent

For our model, both galaxies are assumed to have a stellar mass $M_{\star} =  10^{8.2}$\msun. The dark matter halo mass is adjusted to reflect that our star-forming UDGs have halos typically $60\%$ more massive than non-UDG objects with similar $M_\star$, resulting on $M_{\rm halo} \sim 10^{10.8}$\msun\; and $M_{\rm halo} \sim 10^{10.6}$\msun\; for UDGs and non-UDGs, respectively. We assume an NFW profile with concentration $c=15$, and a scale radii $2.5 ~ kpc$ and $0.9 ~ kpc$ for the UDG and non-UDG respectively (this values where chosen with the average or both population in this stellar mass range). In addition, we considered the fact that at a given stellar mass UDGs have a significantly higher gas fraction, being $M_{\rm gas}/M_{\star} \sim 2$ for a UDG and $0.7$ for a non-UDG. For the other parameters we use $\Delta_{\rm vir} = 200$ and $\rho^0_c = 148 ~ M_{\odot}/kpc^3$. \\ 

\begin{figure*}
\centering
\includegraphics[width=1.0\textwidth]{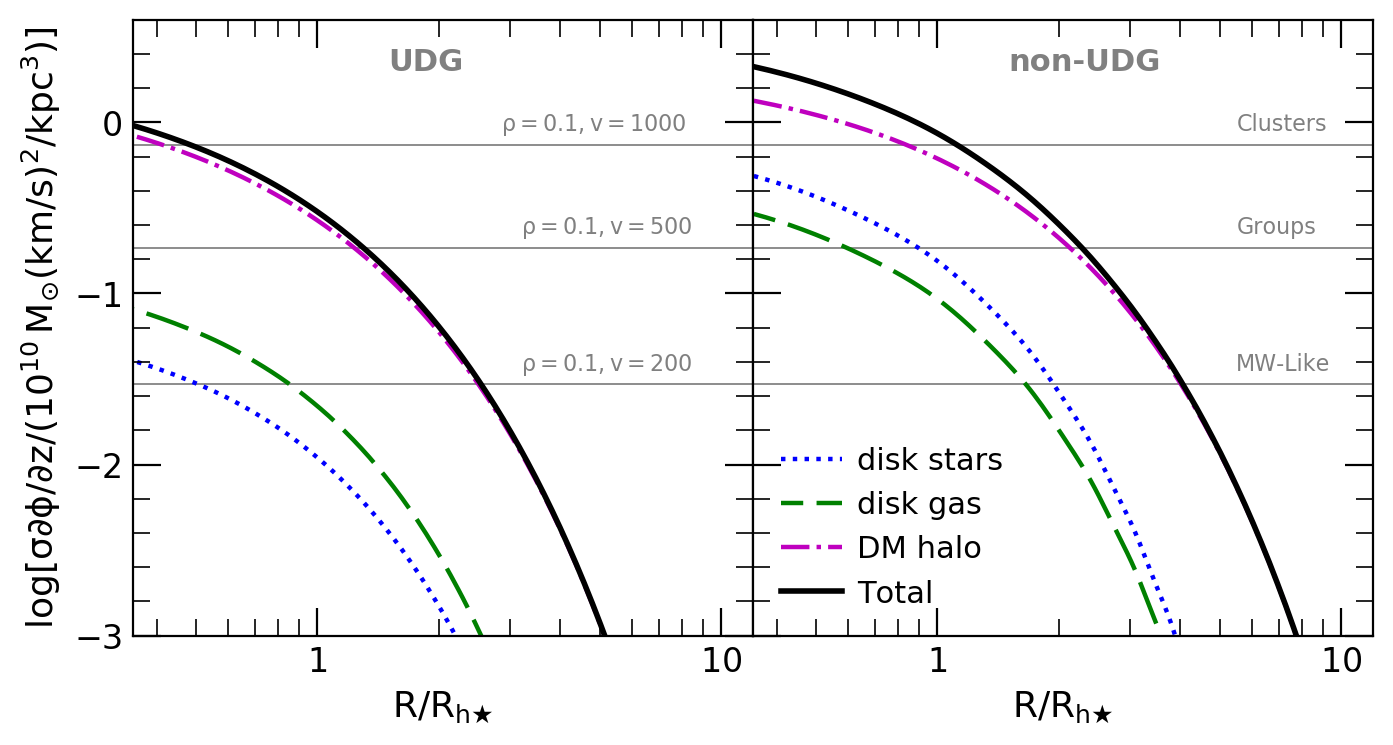}
\caption{Vertical force per unit area as a function of cylindrical radii for two galaxy models: UDG in the left panel and non-UDG in the right. In both cases, we include a stellar (blue dotted line) and gas (green dashed line) exponential disk, embedded in an NFW dark matter halo (magenta dash-dotted line). In both panels, the total contribution (stars+gas+DM) is presented with a black solid line. The grey horizontal lines indicate the ram pressure ($\rho_{ICM} v^2$) for typical values in three different host halo environments: clusters, groups, and galactic halos. All the gas below a given combination of gas density and velocity (gray horizontal lines) would be removed by ram-pressure in this simplified model. Note that the larger mass density in non-UDG galaxies translate on larger restoring forces (y-axis) making them more resilient to ram-pressure stripping.}
\label{fig:ram_pressure}
\end{figure*}

We calculate the values of $z$ at a given $R$, such that the total force is maximal. Thus, in Fig.~\ref{fig:ram_pressure} we present the total vertical force per unit area of the galaxy (UDG in the left panel, non-UDG in the right panel) as a function of cylindrical radius $R$ normalized to the half-mass radius $R_{h \star}$. The different components are shown with different lines: dotted blue for the stellar disk, dashed green for the gas disk, and dotted-dash purple for the dark matter halo. The thick black solid line indicates the total force per unit area resulting in the galaxies, which shall be compared to different ram-pressure values (gray horizontal lines) to determine the radius of gas stripping.\\

As a guideline, we include in Fig.~\ref{fig:ram_pressure} three typical values for ram-pressure in clusters, groups and MW-like halos, corresponding to an ISM density $\rho = 0.1\; \rm M_\odot/\rm kpc^{3}$ and velocities $1000$, $500$ and $200$ km/s, respectively. In this model, the gas disk gets stripped from the disk anywhere where the ram-pressure force is larger than the galaxy self-gravity. Note that due to the dependence of the restoring force on the scale radius of the disk, the UDG would consistently get the gas disk truncated at a smaller radius than the non-UDG. For example, in the intermediate case shown here, with $\rho=0.1$ and $v=500$ km/s, the UDG would loose all gas beyond $R/R_{h,\star}\sim 1.2$ while the non-UDG case would only get truncated beyond $R/R_{h,\star}\sim 2$. The situation is even more clear for higher-density environments like a cluster. For our most extreme ram-pressure case, the UDG gets stripped of almost all of its disk, while the non-UDG would be able to retain all its gas within $\sim R_{h,\star}$. These results lends support to the claim that enhanced ram-pressure stripping in the extended UDGs translates into shorter timescales for quenching, as found in our simulated sample (see Fig.~\ref{fig:quech_timescales}).\\

\label{lastpage}

\end{document}